\documentclass[aps,prb,reprint,amsmath,amssymb,groupedaddress]{revtex4-1}

\usepackage[colorlinks=true, allcolors=blue]{hyperref}
\usepackage{graphicx}
\usepackage{dcolumn}
\usepackage{bm}
\usepackage{xcolor}
\usepackage{color}
\usepackage{siunitx}
\usepackage{caption} 
\usepackage{subcaption}
\usepackage{todonotes}
\usepackage{threeparttable}
\usepackage{gensymb}
\usepackage{float}
\usepackage{soul}
\usepackage{xr}
\usepackage{ulem}
\usepackage{braket}
\usepackage{amsmath}
\usepackage{relsize}
\usepackage{bbold}
\usepackage[export]{adjustbox}

\begin{document}

\title{Impact of structural distortions on the correlated electronic structure of orbital-selective Mott insulating Na$_3$Co$_2$SbO$_6$ under strains}

\author{Nam Nguyen$^{1}$, Alex Taekyung Lee$^{2,3}$, Anh T.\ Ngo$^{2,3}$, and Hyowon Park$^{1,2}$}

\affiliation{$^1$Department of Physics, University of Illinois at Chicago, IL 60607, USA \\
$^2$Materials Science Division, Argonne National Laboratory, IL 60439, USA \\
$^3$Department of Chemical Engineering, University of Illinois at Chicago, IL 60607, USA
}

\date{\today}

\begin{abstract}
Na$_{3}$Co$_{2}$SbO$_6$ is a promising candidate to realize the Kitaev spin liquid phase since the large Kitaev spin exchange interaction is tunable via the change in electronic structure, such as the trigonal crystal field splitting ($\Delta_{TCF}$). Here, we show that the uncorrelated electronic structure of Na$_{3}$Co$_{2}$SbO$_6$ is rather insensitive to the strain effect due to the low crystal symmetry accompanied by oxygen displacements and the presence of Sb $s$ orbitals. Using density functional theory plus dynamical mean field theory, we find that the correlated electronic structure of Na$_{3}$Co$_{2}$SbO$_6$ is an orbital selective Mott insulating state where the trigonal $a_{1g}$ orbital is insulating due to the nearly full occupation, while other $d$ orbitals behave as typical Mott insulators, resulting in tunability of $\Delta_{TCF}$ under the strain effect effectively. The sign change of $\Delta_{TCF}$ can occur as the in-plane tensile strain is applied, and the Kitaev spin liquid phase could be possibly realized due to the strongly suppressed $\Delta_{TCF}$ under tensile strain. Our results show that the local Co-site symmetry and dynamical correlation effects will play an important role in engineering the novel magnetic phase in this and related materials.

\end{abstract}

\maketitle


Bond-dependent spin exchange interaction in a honeycomb structure plays a crucial role in materials for stabilizing the Kitaev quantum spin liquid (QSL) phase~\cite{Takagi2019KitaevQS}.
In particular, 4$d$ or 5$d$ orbitals in a honeycomb structure have been attracting much interest in realizing anisotropic spin interactions due to their large spin-orbit coupling (SOC)~\cite{PhysRevLett.105.027204}. However, the spatially extended 4$d$ or 5$d$ orbitals typically generate longer-range spin interaction terms, making the magnetic phase diagram much more complex. Moreover, it is rather difficult to tune the SOC effect via structural changes such as strain or pressure. Recently, 3$d$ orbitals have been suggested to stabilize the Kitaev QSL phase due to the anisotropic hopping mechanism despite their small SOC~\cite{QSL_Condition_1,QSL_Condition_2}. In this regard, Na$_{3}$Co$_{2}$SbO$_6$ (NCSO) has drawn much attention as a possible material candidate due to its tunability close to the QSL phase~\cite{PhysRevLett.125.047201}. 

NCSO has monoclinic structure with space group $C2/m$ and undergoes a magnetic phase transition from a paramagnetic (PM) insulating 
state to an antiferromagnetic (AFM) insulating state at $T_N\sim 5$ K \cite{VICIU,WONG,POLITAEV,PhysRevMaterials.3.074405}. The Co$^{2+}$ $(d^7)$ ion is in the high spin (HS) state with the electronic configuration of $t_{2g\uparrow}^{3}t_{2g\downarrow}^{2}e_{g\uparrow}^{2}$ $(S=3/2, L=1)$, forming a spin-orbit-entangled pseudospin $J_{eff}=1/2$ state, from which the Kitaev interaction arises via the second or higher order perturbation \cite{QSL_Condition_1,QSL_Condition_2,PhysRevLett.125.047201, PhysRevB.107.054420,PhysRevB.109.L180413}. 
It is shown that the $t_{2g}$-$e_g$ hopping channel in NCSO enhances the Kitaev interaction $J_K$ while the non-Kitaev terms (Heisenberg $J_H$, off-diagonal $\Gamma$, and trigonal $\Gamma'$) are much weakened~\cite{PhysRevLett.125.047201}. Also, the stability of the QSL phase depends sensitively on the size of the trigonal crystal field splitting ($\Delta_{TCF})$ since a large $\Delta_{TCF}$ can weaken the Kitaev interaction. Therefore, tuning the $\Delta_{TCF}$ or related material parameters via strain or pressure is a priority of driving the system into Kitaev QSL phase.

The experimental estimates of electronic structure in NCSO, such as the sign and magnitude of the $\Delta_{TCF}$, have been relied on fitting the crystal field multiplet calculations to the spectroscopic measurements including X-ray absorption spectroscopy (XAS), inelastic neutron scattering (INS), X-ray linear dichroism (XLD), and X-ray magnetic circular dichroism (XMCD). Kim $et$ $al$ found that the magnitude of the $\Delta_{TCF}$ is 25.1 meV by fitting to the XAS/XLD data, and the $a_{1g}$ orbital is located below the $e_{g}^{\pi}$ orbitals~\cite{XAS_TCF}. 
However, Veenendaal $et$ $al$ reported that the $a_{1g}$ orbital is located above the $e_{g}^{\pi}$ orbitals with $\Delta_{TCF}$ of 35-60 meV by fitting configuration interaction calculations to the XLD and XMCD data~\cite{PhysRevB.107.214443}.
This sign of $\Delta_{TCF}$ has been attributed to the hybridization of $e_{g}^{\pi}$ orbitals with the positive $Sb^{5+}$ ion located in the honeycomb Co layers, and it is also consistent with the analysis using the INS data~\cite{INS_TCF}.

Although most theoretical works of studying the QSL phase in NCSO have focused on computing the anisotropic spin exchange parameters of the localized spin Hamiltonian based on perturbation theory assuming the pure trigonal distortion, more realistic treatment of structural changes in the monoclinic NCSO under strain or pressure and its impact on the correlated electronic structure and the spin-exchange interactions have not been explored yet. Moreover, a Jahn-Teller (JT) distortion is typically allowed in the monoclinic $C2/m$ phase such as the NCSO case,
while other QSL candidate materials such as BaCo$_{2}$(AsO$_4$)$_2$(BCAO)~\cite{PhysRevB.107.054420,PhysRevB.108.064433,10Dq}, $\gamma$-BaCo$_{2}$(PO$_4$)$_2$~\cite{PhysRevB.97.134409,PhysRevMaterials.7.024402}, or $\alpha$-RuCl$_3$~\cite{nano14010009}, have a pure trigonal distortion associated with 
$R\bar{3}$. A JT distortion can further split degenerate orbitals into different energy levels, 
which could result in some novel physical phenomena~\cite{JT-ice}. In addition, the properties of material parameters such as $\Delta_{TCF}$ depend on the orbital-dependent correlation effect of the Co$^{2+}$ ion due to the strong on-site $U$ and its hybridization effect with the ligand fields of O$^{2-}$ and Sb$^{5+}$ orbitals, which can be treated accurately using first principles.

Here, we study the impact of structural distortions of NCSO under strain on its correlated electronic structure and spin-exchange interactions using first-principles. We first perform structural relaxations of strained structures using density functional theory (DFT)+$U$~\cite{LDA+U} based on the projected-augmented wave (PAW) method~\cite{PAW} as implemented in the Vienna \textit{ab initio} simulation package (VASP)~\cite{vasp1,vasp2}. All atomic positions of NCSO are relaxed to obtain the strained structures, while the volume is fixed to 270.84 \AA, which is the zero-strain relaxed volume. The exchange-correlation energy functional was treated using the generalized gradient approximation (GGA) adopting the Perdew-Burke-Ernzerhof (PBE) functional~\cite{PBE}. 
The cutoff energy for the plane-wave basis was used as 600 eV with the Gamma-centered 8$\times$4$\times$8 $k-$point mesh. The Hellmann-Feynman force on each atom was set to be smaller than 0.01 eV/\AA \: for convergence. We imposed the Hubbard $U$ and Hund coupling $J$ on the Co $3d$ orbitals with $U$=5 eV and $J$=0.8 eV for DFT+$U$ calculations.
We also impose the N\'{e}el AFM order for the the relaxation since this is the ground-state spin configuration (see Appendix A).

We then performed DFT+dynamical mean field theory (DMFT) calculation~\cite{SINGH2021107778} for each relaxed structure to study the dynamical correlation effects on NCSO. 
For DMFT calculations, we construct the Wannier Hamiltonian using the Wannier90 code~\cite{Wannier90} for the Co $d$, O $p$, and Sb $s$ orbitals.
The Wannier Hamiltonian of this $spd$-model and its comparison to the DFT band is given in Appendix B and C. Due to the low point group symmetry ($C_{2h}$) of CoO$_6$ octahedra, the off-diagonal elements in 
Wannier Hamiltonian are non-negligible.
To minimize the effect of the off-diagonal elements, we diagonalize the Co $d$ block 
in the full $spd-$Hamiltonian by applying unitary rotation matrix \cite{LeePRB2023}, 
and use the diagonalized Hamiltonian for DFT+DMFT loop. 
Then, we solve the DMFT impurity problem using the continuous time quantum Monte Carlo method~\cite{CTQMC4} with the Hubbard $U$=5 eV and the Hund's coupling $J$=0.8 eV to treat correlations of Co $3d$ orbitals.
Once the DMFT calculations converge, all final results including the spectral function, the occupancy matrix, and the self-energy $\Sigma (\omega)$ are obtained as the original trigonal basis using the unitary transform. Here, we adopt single-site DMFT to treat the local dynamical correlation effect since this approximation has been successfully applied to the $3d$ transition-metal compounds in a similar honeycomb structure~\cite{Single-Site-DMFT,Samanta2024-pi}. The details of DFT+DMFT calculation are given in Appendix D.

\begin{figure}[htb!]
  \centering
  \includegraphics[width=0.20\textwidth]{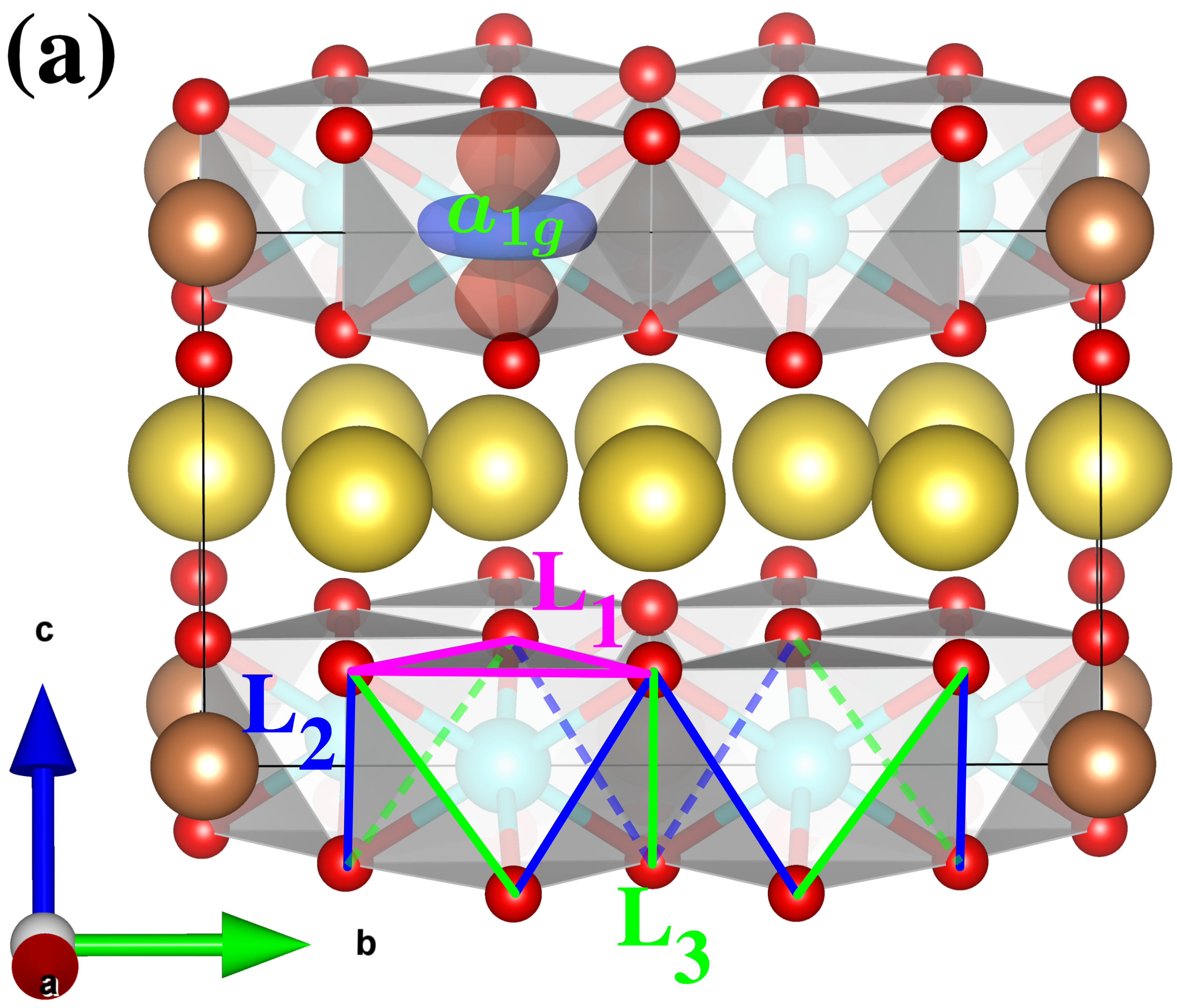}
  \includegraphics[width=0.275\textwidth]{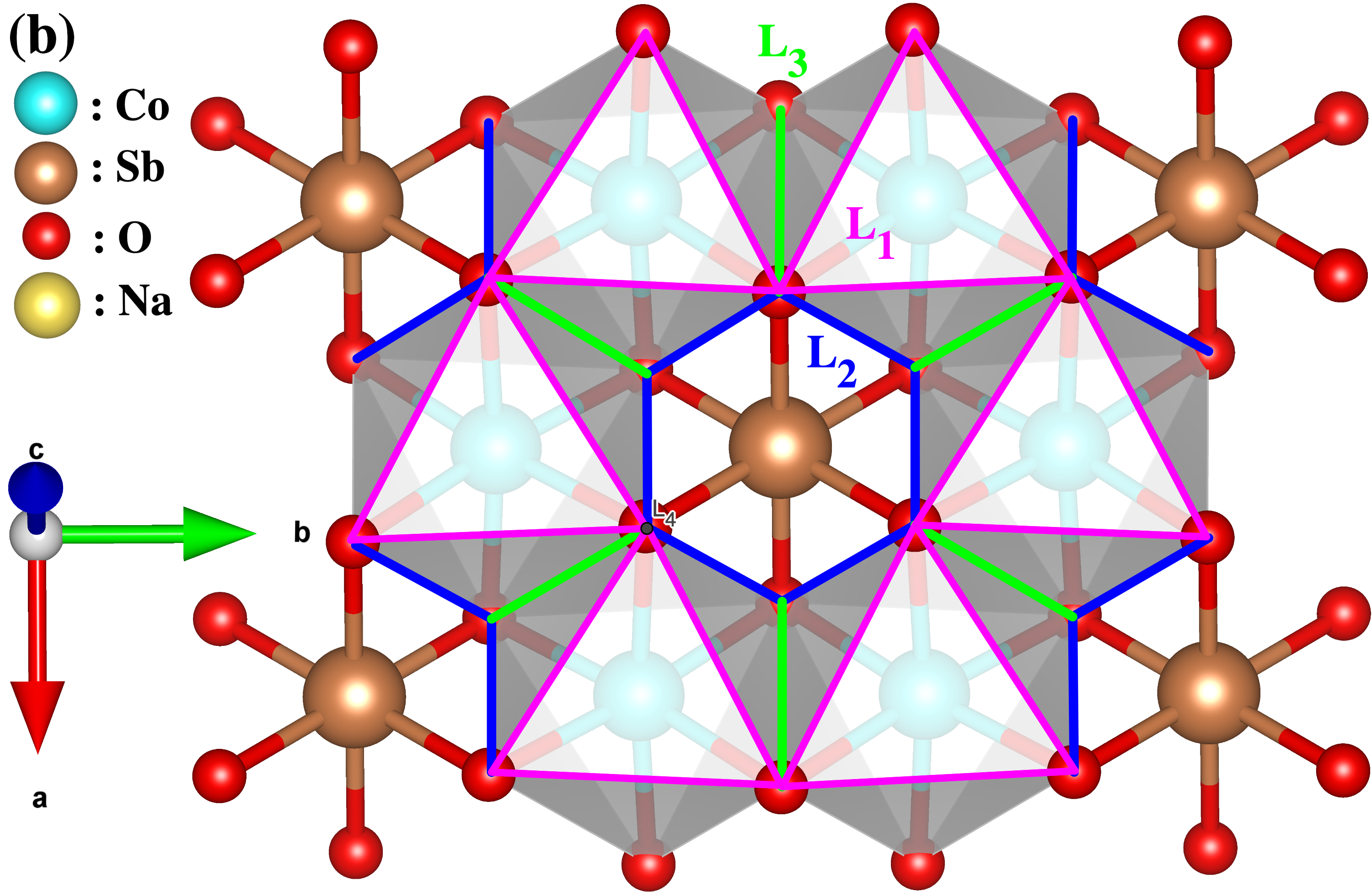}
  \\
  \includegraphics[width=0.48\textwidth]{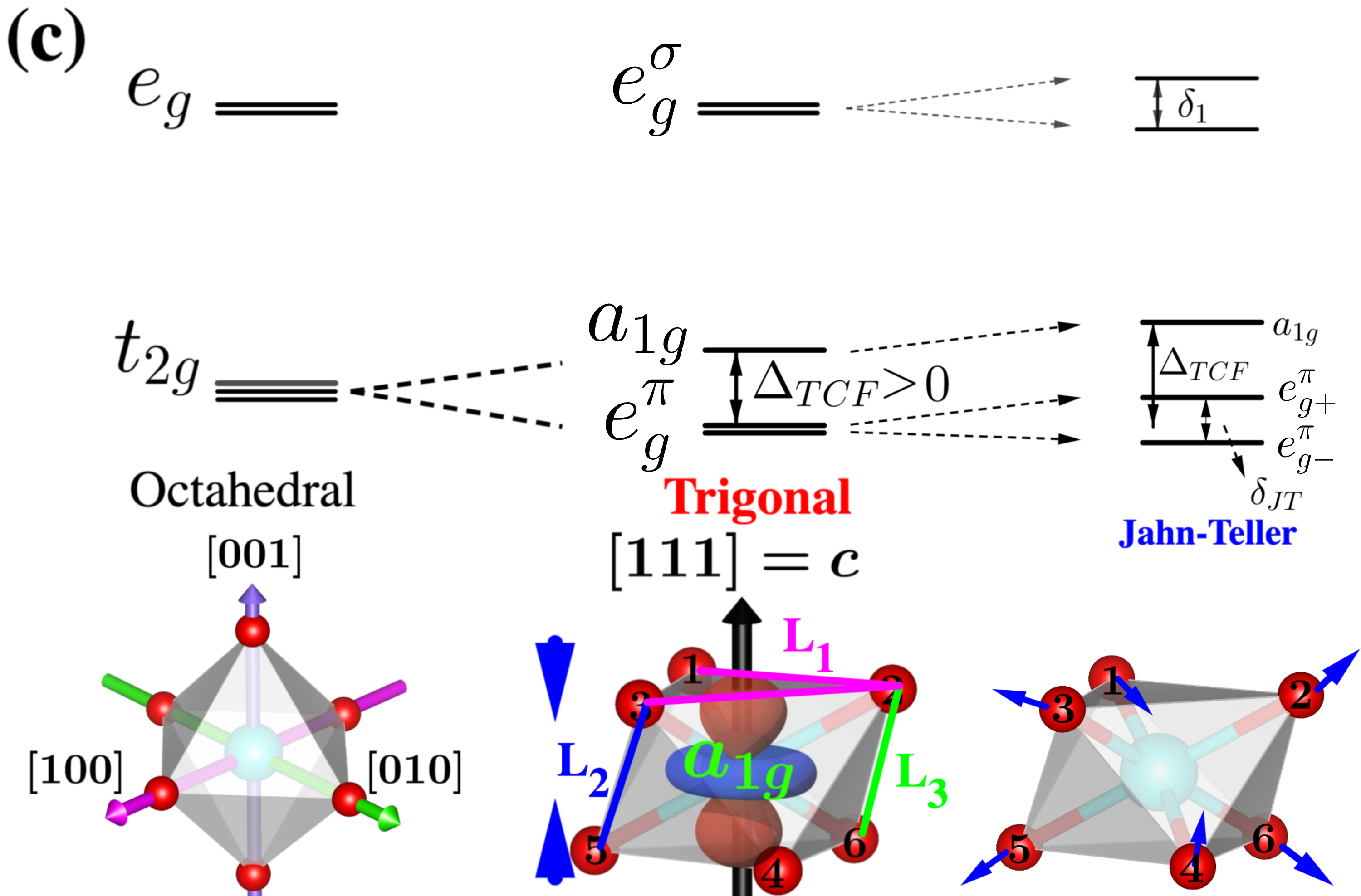}
\caption{The (a) side and (b) top views of the crystal structure, Na$_3$Co$_2$SbO$_6$. $L_1\neq L_2\neq L_3$ due to structural distortions. (c) A schematic diagram of orbital energy level changes due to the trigonal compression and oxygen displacements (The $c-$axis corresponds to the [111] direction along the local Co-O axis).}
  \label{Structure}
\end{figure}

Experimental studies of neutron and X-ray diffractions show that NCSO forms a centered monoclinic cell with the space group $C2/m$ ([No.\:12])~\cite{VICIU,WONG,POLITAEV,PhysRevMaterials.3.074405}. The DFT+U relaxed structure with the AFM N\'{e}el configuration exhibits a slightly larger volume along with the elongated in-plane $b-$axis compared to the experimental one (see Table\:\ref{structural parameters}). While Co ions in NCSO form a layered honeycomb structure, the local Co-O octahedra follow the trigonal symmetry (Fig.\:\ref{Structure} (a,b)), similarly to those in transition-metal dihalides such as CoI$_2$ with the space group $P\bar{3}m1$ ([No.\:164]). This trigonal distortion splits the $t_{2g}$ triplet of the Co 3$d$ manifold into an $a_{1g}$ singlet and a degenerate $e_{g}^{\pi}$ doublet, where the trigonal crystal field splitting $\Delta_{TCF}$ is defined as $E_{a_{1g}}-E_{e_g^{\pi}}$ (Fig.\:\ref{Structure}c)~\cite{PhysRevB.105.245153, Orbital_Shape, Winter_2022,JT-ice}. 
However, the point group symmetry of the local Co-site in NCSO is $C_{2h}$, which is lower than that of CoI$_2$ ($D_{3d}$). This lower local-site symmetry can play an important role in tuning the correlation effects in NCSO and related low-symmetry compounds.

\begin{table*}[htb!]
\caption{Lattice and structural parameters of Na$_3$Co$_2$SbO$_6$ without any strains ($\Delta a=0$).} 
\begin{tabular}{|c| c| c| c| c| c| c| c| c| c| c}
\hline
Na$_3$Co$_2$SbO$_6$  & $V$ [\AA$^3$]  & $a$ [\AA] & $b$ [\AA] & $c$ [\AA] & $\beta [^0]$ & $L_1$ [\AA] & $L_2$ [\AA] & $L_3$ [\AA] &Co-O [\AA]

  \\ [0.5ex] 
\hline
DFT+U N\'{e}el & 270.84 & 5.40 & 9.33 & 5.66 &108.5 &	3.212/3.207/3.199
&2.828/2.809/2.809
 & 3.019/2.999/2.999
 &2.161/2.166
\\
Viciu $et$ $al.$\cite{VICIU} & 267.22 & 5.37
& 9.28 & 5.65
& 108.5 
& 3.172/3.168/3.157 & 2.757/2.785/2.785 & 2.915/2.927/2.927 &2.136/2.125
\\
Wong $et$ $al.$\cite{WONG} &266.81 & 5.36
& 9.29 & 5.65
& 108.4 
& 3.169/3.167/3.152 & 2.758/2.784/2.784&2.915/2.926/2.926 & 2.132/2.127
\\
\hline
\end{tabular}
\label{structural parameters}
\end{table*}

The trigonal distortion of the Co-O octahedra can be parametrized by the $L_2/L_1$ and $L_3/L_1$ ratios where $L_1$ is the in-plane O-O bond lengths and $L_2$ and $L_3$ are the out-of-plane O-O bond lengths as shown in Fig.\:\ref{Structure}(a,b). Here, $L_3$ is the edge that intersects the nearest-neighbor Co-Co bonding, while $L_2$ is not. The values of $L_1$, $L_2$, and $L_3$ computed for the fully relaxed NCSO structure without any strains are given and compared to known experimental structures in Table \ref{structural parameters}. Under zero strain ($\Delta a$=0), the fully relaxed NCSO structure shows the trigonal compression along the $c-$axis, yielding  the average $L_2/L_1$=0.878 and $L_3/L_1$=0.937, both of which are less than one. Additionally, the monoclinic distortion in NCSO results in unequal Co–O bond lengths within the Co–O octahedra, akin to the JT distortion observed in various perovskite transition-metal oxides. This is due to the anisotropic elongation of the oxygen positions, which alters the ratio between the Co–O bond lengths along the $c$-axis and those in the $ab$-plane, 
i.e., (Co-O)$_{\hat{z}}$/(Co-O)$_{\hat{x}/\hat{y}}$
Although the JT distortion can be small in the unstrained state, its effects become more pronounced 
under applied strain, and a significant JT distortion is also evident 
in the experimental structure~\cite{VICIU,WONG}.

Without the JT distortion, the trigonal $a_{1g}$ and $e_{g}^{\pi}$ orbitals are the appropriate 
basis functions for diagonalizing the local Co $d$ Hamiltonian under trigonal distortion (see Appendix B). In general, even in the absence of JT distortions, unequal values of $L_2$ and $L_3$ ($L_2\neq L_3$) 
 are allowed under trigonal symmetry. 
For example, BCAO with the space group $R\bar{3}$ ([No.\:148]) shows $L_2\neq L_3$ with the 
local-site symmetry of $C_3$, but there are no JT distortions and Co-O bonds are still equivalent. In contrast, all out-of-plane O-O bond lengths are equivalent with $L_2$=$L_3$ in CoI$_2$. In NCSO, however, the local-site symmetry is further reduced by the JT distortion, 
which induces $L_2\neq L_3$ and an energy splitting between the $e_g^{\pi}$ orbitals ($\delta_{JT}$). 
This results in the non-zero off-diagonal terms in the local Co $d$ Hamiltonian represented using the trigonal basis.

To understand the response of $\Delta_{TCF}$ and $\delta_{JT}$ due to structural changes, we compute the local Co-site $spd-$Hamiltonian by projecting the DFT band structure to trigonal basis functions.
In the case of both CoI$_2$ and BCAO, the off-diagonal terms of the local Hamiltonian in the $t_{2g}$ manifold represented using trigonal basis functions are zeros, and $\Delta_{TCF}$ can be defined as $E_{a_{1g}}-E_{e_{g}^{\pi}}$. The $\delta_{JT}$ term is zero, thus the two $e_{g}^{\pi}$ orbitals are degenerate.
For NCSO, the low Co-site symmetry lifts the degenerate $e_{g}^{\pi}$ orbital energy levels and results in non-negligible off-diagonal terms in the trigonal Hamiltonian. 
We define both terms as $\Delta_{TCF}=E_{a_{1g}}-(E_{e_{g+}^{\pi}}+E_{e_{g-}^{\pi}})/2$ and $\delta_{JT}=E_{e_{g+}^{\pi}}-E_{e_{g-}^{\pi}}$.

\begin{figure}[htb!]
  \centering
 \includegraphics[width=0.40\textwidth]{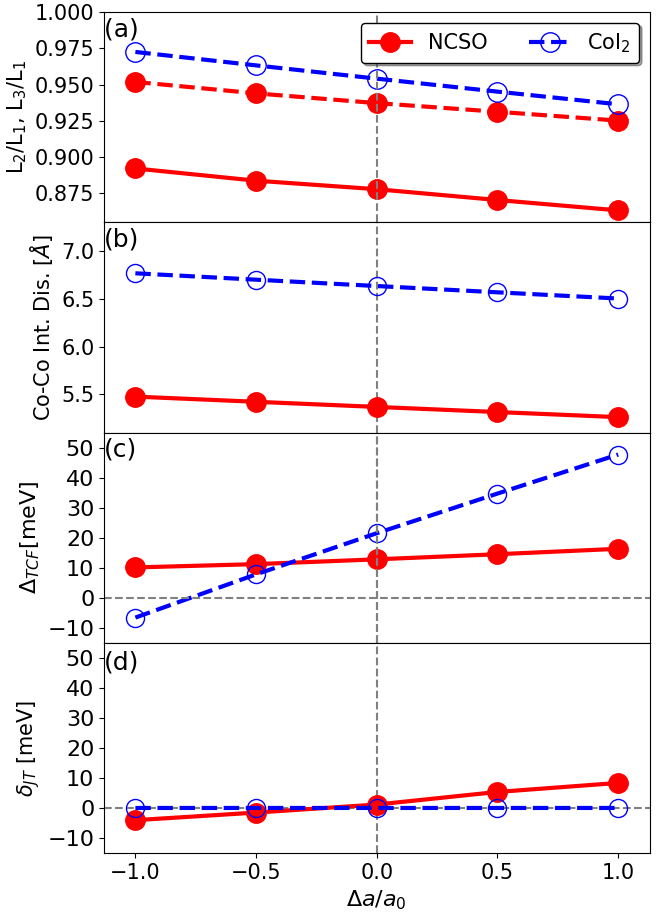}
\caption{The strain effect on structural parameters of Na$_3$Co$_2$SbO$_6$ and CoI$_2$: (a) the ratio $L_2/L_1$ (solid line) and $L_3/L_1$ (dashed line), (b) the Co-Co interlayer distance, (c) the trigonal crystal field $\Delta_{TCF}$, and (d) the energy splitting within the $e_{g}^{\pi}$ orbitals ($\delta_{JT}$) }
  \label{spd_Strain}
\end{figure}

Fig.\:\ref{spd_Strain}a shows that the trigonal distortions of NCSO and CoI$_2$ occur even at zero strain,
resulting in a Co–O octahedron compressed along the $c-$axis with $L_2/L_1$ and $L_3/L_1$ $<$1.
The corresponding structural values agree well with the experimental data (see Table \ref{structural parameters}) and
these distortions are further enhanced under tensile strains ($\Delta a>0$). In Fig.\:\ref{spd_Strain}c, we plot $\Delta_{TCF}$ as a function of strain for each relaxed structure 
by extracting the on-site orbital energies from trigonal Wannier orbitals projected 
onto the uncorrelated DFT band structure. In both compounds, the $\Delta_{TCF}$ is positive ($\sim$21.5meV for CoI$_2$ and $\sim$12.8meV for NCSO) at zero strain. 
This $\Delta_{TCF}$ of NCSO is consistent with the experimental estimate obtained using INS~\cite{INS_TCF}.
Under the compression along the trigonal $c-$axis,
the negative $\Delta_{TCF}$ is conventionally expected due to the stronger hybridizaton of $e_{g}^{\pi}$ orbitals with the O ions when $L_2/L_1<1$~\cite{Winter_2022} as the geometric lobes of the $e_{g}^{\pi}$ ($a_{1g}$) orbitals point to the octahedral faces parallel (perpendicular) to the $c-$axis \cite{Orbital_Shape,PhysRevB.105.245153,PhysRevB.104.014414}. However, it is important to note that the Wannier orbitals are maximally localized, and the hybridization effect on the crystal field splitting is minimized. 
Moreover, the Sb $s$ orbitals also hybridize with O $p$ orbitals and weaken the Co-O hybridization effect. 
We find that the Co-Co interlayer distance does not have much effect on the crystal field splitting as it is large and does not depend much on strains (Fig.\:\ref{spd_Strain}b).

Although the structural changes ($L_2/L_1$ and $L_3/L_1$) due to strains are similar for both NCSO and CoI$_2$ (Fig.\:\ref{spd_Strain}a), the response of the material parameters including $\Delta_{TCF}$ and $\delta_{JT}$ of NCSO are much less sensitive to strains. 
The $\Delta_{TCF}$ of CoI$_2$ increases under tensile strains and changes sensitively about 27meV per 1\% of strain (Fig.\:\ref{spd_Strain}c).
This trend of the positive $\Delta_{TCF}$ is consistent with that of TiCl$_2$ (the same point group as CoI$_2$) when $L_2/L_1<1$ as discussed in Georgescu $et$ $al$~\cite{PhysRevB.105.245153}. 
For NCSO, the change of structural parameters does not directly impact the energy levels of localized $\{a_{1g},e_g^{\pi}\}$ orbitals due to the low Co-site symmetry. At zero strain, the JT distortion further splits the $e_{g}^{\pi}$ orbitals of NCSO with the crystal splitting of $\delta_{JT}\simeq 1.1$meV and $\delta_{JT}$ changes about 6meV per 1\% of strain (Fig.\:\ref{spd_Strain}d).

\begin{figure}[htb!]
   \centering
   \includegraphics[width=0.42\textwidth]{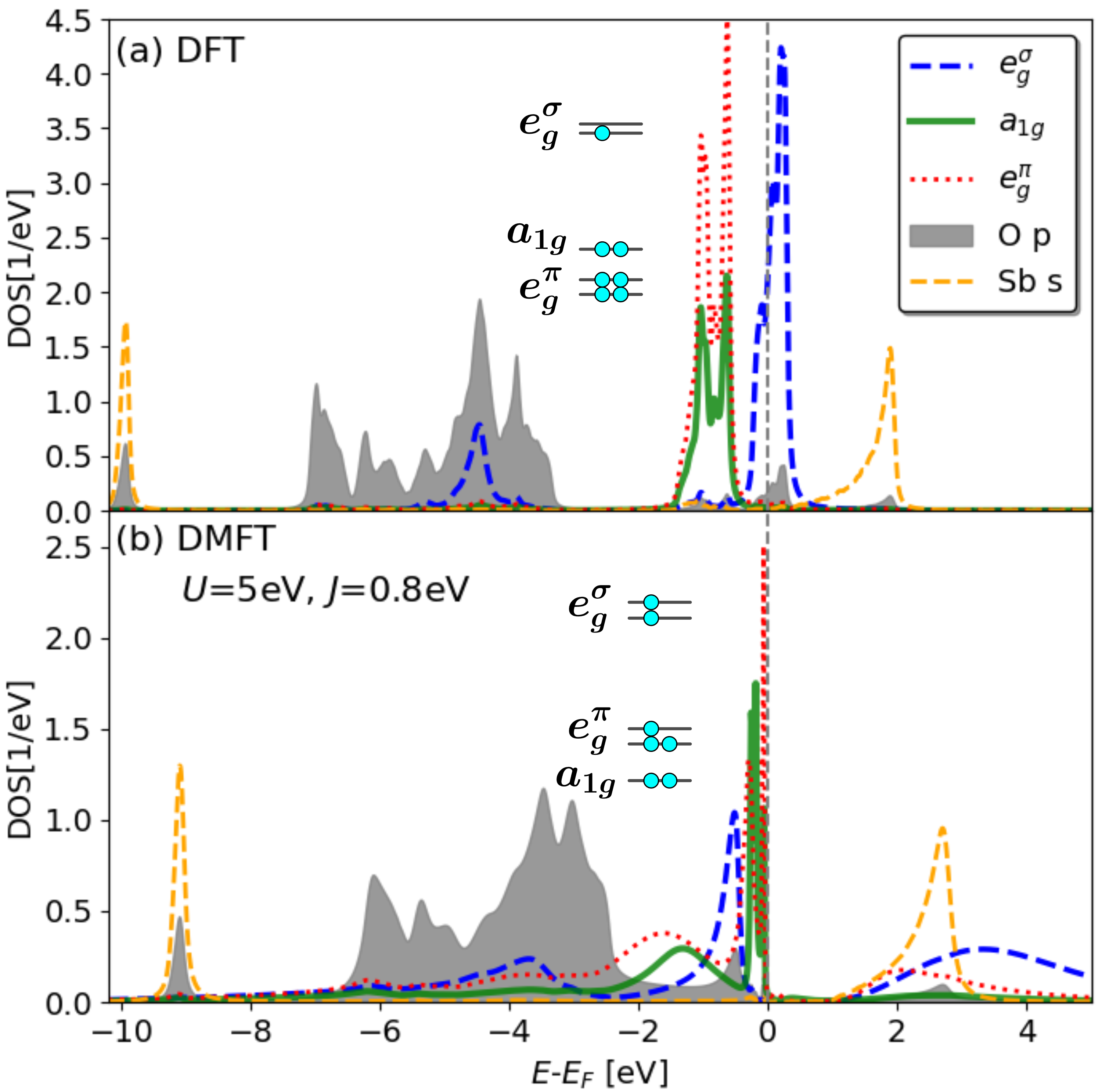}

   \hspace{2cm}
   
   \includegraphics[width=0.44\textwidth]{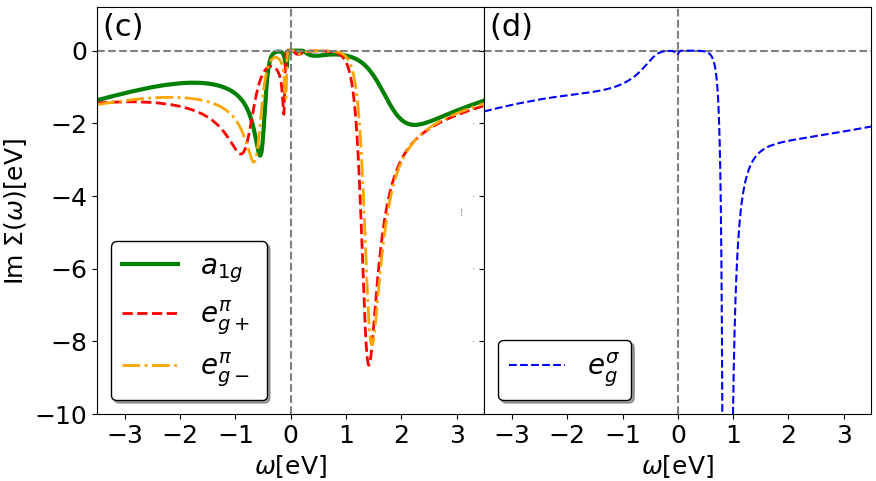}
   
   \caption{(a) DFT DOS of NCSO at zero strain. (b) DFT+DMFT DOS of NCSO with $U$=5eV and $J$=0.8eV within paramagnetic phase. The imaginary part of the DMFT self-energy $\Sigma(\omega)$ obtained for (c) $a_{1g}$, $e_{g}^{\pi}$, and (d) $e_g^{\sigma}$ orbitals.} 
   \label{0_DOS_Self_E}
\end{figure}

We now study the effect of dynamical correlations on the electronic structure of strained NCSO in the paramagnetic phase. Since the five Co $d$ orbitals in NCSO are not equivalent, we treat the correlation effect of them separately with distinct DMFT self-energies. Without the effects of the Hubbard $U$ and the Hund's coupling $J$ (DFT calculation), the ground state of NCSO is the metallic state with a low-spin electronic configuration $(e_{g}^{\pi})^4(a_{1g})^2(e_g^{\sigma})^1$. The DFT DOS shows that the low-energy bands of $a_{1g}$ and $e_{g}^{\pi}$ orbitals are fully filled, leaving a single electron on the doublet $e_g^{\sigma}$ state (Fig.\:\ref{0_DOS_Self_E}a).

The paramagnetic DFT+DMFT calculation using $U=5$eV and $J=0.8$eV shows that NCSO at the zero strain becomes an orbital-selective Mott insulator with an energy gap of $1$ eV.
The $e_g^{\sigma}$ bands under the effect of strong correlations become almost half-filled (Fig.\:\ref{0_DOS_Self_E}b) with the occupancy $N_{e_g^{\sigma}}=2.16$. Fig.\:\ref{0_DOS_Self_E}d shows that the imaginary part of self-energy ($Im\Sigma(\omega)$) of the $e_g^{\sigma}$ orbitals has a strong pole (divergent) at $\omega\sim 0.8$ eV with the broad upper Hubbard bands near 3.5eV above the Fermi energy, indicating that the $e_{g}^{\sigma}$ orbitals exhibit the typical strongly-correlated Mott insulating behavior.
We find that this Mott state is robust against the strain effect.

Unlike the half-filled $e_g^{\sigma}$ orbitals, the $a_{1g}$ and $e_{g}^{\pi}$ orbitals are partially filled with one shared hole in the $t_{2g}$ manifold ($h_{t_{2g}}=1.0$), implying that the correlation effect can arise due to the multi-orbital physics in a non-trivial way. 
The partially filled occupancies among the $a_{1g}$ and $e_{g}^{\pi}$ orbitals are due to the large Hund's coupling $J$ acting on the $t_{2g}$ manifold with a small $\Delta_{TCF}$.
This shared hole is mostly located on the $e_{g}^{\pi}$ orbitals where the broad upper Hubbard band is formed near 2eV above the Fermi energy.
The $Im\Sigma(\omega)$ pole of the $e_{g}^{\pi}$ orbital is located near 1.5eV above the Fermi energy and the pole strength is weaker than the $e_g^{\sigma}$ orbital one (Fig.\:\ref{0_DOS_Self_E}c).
This means that the Mott insulating state of the $e_{g}^{\pi}$ orbital shows moderate correlations, which could be tunable by strains. 
The self-energies for $e_{g+}^{\pi}$ and $e_{g-}^{\pi}$ orbitals are similar despite the small JT splitting between two orbitals.

The $Im\Sigma(\omega)$ for the $a_{1g}$ orbital shows much broader peaks with an almost filled $a_{1g}$ occupancy ($N_{a_{1g}}=1.8$).
The peak position is located near 2.0eV above the Fermi energy, which corresponds to the Sb $s$ orbital energy level due to their hybridization (Fig.\:\ref{0_DOS_Self_E}c).
This shows that the $a_{1g}$ orbital is close to a hybridization-induced insulating state rather than a Mott insulating state.
Our result also suggests that the one-particle energy level of the $a_{1g}$ orbital can be effectively shifted lower due to the nearly full occupation, resulting in the negative trigonal crystal field splitting ($\Delta_{TCF}<0$), while the $a_{1g}$ orbital energy is higher than the $e_{g}^{\pi}$ orbital ones at the DFT level ($\Delta_{TCF}>0$). We want to emphasize that the orbital-dependent correlation effect plays an important role in tuning the effective Co $3d$ orbital levels and the resulting $\Delta_{TCF}$ splitting under strains. This is also in sharp contrast to the CoI$_2$ case where the DMFT occupancy of the $a_{1g}$ orbital is smaller than the $e_{g}^{\pi}$ one ($\Delta_{TCF}>0$) due to the absence of the Sb $s$ orbital. 

\begin{figure}[htb!] 
   \centering
   \includegraphics[width=0.40\textwidth]{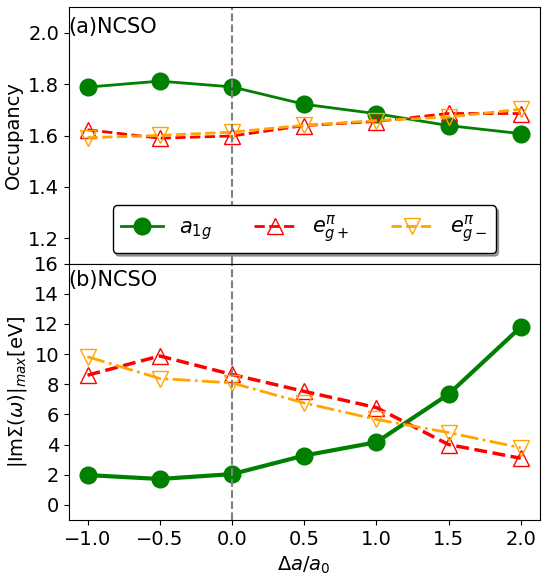}

   \hspace{1mm}

   \includegraphics[width=0.40\textwidth]{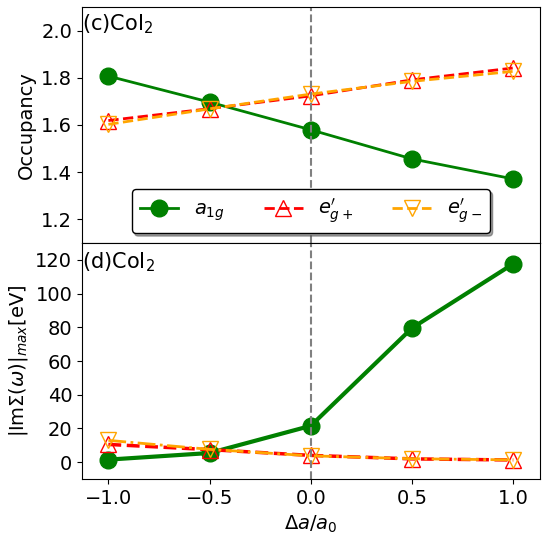}

\caption{DFT+DMFT results of (a) the occupancy and (b) the pole strength $|Im\Sigma(\omega)|_{max}$  of the $a_{1g}$, $e_{g+}^{\pi}$, and $e_{g-}^{\pi}$ orbitals in NCSO as a function of strain. (c) The occupancy and (d) the pole strength $|Im\Sigma(\omega)|_{max}$  of the $a_{1g}$, $e_{g+}^{\pi}$, and $e_{g-}^{\pi}$ orbitals in CoI$_2$ for comparison.} 
   \label{Occupancy NCSO vs CoI2}
\end{figure}

Now, we study the strain effect on the change of correlations in NCSO and CoI$_2$.
Fig.\:\ref{Occupancy NCSO vs CoI2}a shows that the DMFT occupancies in NCSO are rather insensitive to compressive strains as the $a_{1g}$ occupancy is close to the full occupation ($N_{a_{1g}}\sim 1.8$).
Under tensile strain, $N_{a_{1g}}$ gradually decreases, and the $a_{1g}$ and $e_{g}^{\pi}$ occupancies are almost degenerate at $\Delta a/a_0\sim 1.2\%$. We expect that the sign change of $\Delta_{TCF}$ could occur near the tensile strain $\Delta a/a_0\simeq 1.2\%$ and possibly realize the spin liquid phase due to the strongly suppressed $\Delta_{TCF}$. The change of the DMFT occupancy under tensile strain is closely related to the orbital-dependent correlation effect
since the occupancy increases (decreases) gradually as the $|Im\Sigma(\omega)|_{max}$ decreases (increases) (Fig.\:\ref{Occupancy NCSO vs CoI2}b). For the CoI$_2$ case, the trend of the occupancy change is more rapid (Fig.\:\ref{Occupancy NCSO vs CoI2}c) but still consistent with the change of $\Delta_{TCF}$ as the $a_{1g}$ and $e_{g}^{\pi}$ occupancies are almost degenerate at $\Delta a/a_0\sim -0.5\%$ where $\Delta_{TCF}$ becomes nearly zero. 
Overall, the tensile (compressive) strain favors the stronger correlation effect of $a_{1g}$ ($e_{g}^{\pi}$) orbital as it is less hybridized with O $p$ orbitals. 
The difference between two compounds originates from the low Co-site symmetry and the presence of Sb $s$ orbitals in NCSO.


In summary, we studied the strain and structural distortion effects on electronic and magnetic properties of NCSO using first-principles calculations.
We found that the orbital-dependent correlation effect of Co ions will play an important role in tuning the material parameters under strains.
While the $a_{1g}$ orbitals are insulating due to the nearly full occupaction, other Co $d$ orbitals are more strongly correlated as a Mott insulating state. 
The electronic structure dependence on strains is much weaker for NCSO compared to the CoI$_2$ case due to its low Co-site symmetry originated from the O distortion and the presence of the Sb $s$ orbitals.
The study of spin-exchange interactions incorporating such realistic and correlated electronic structures~\cite{PhysRevB.61.8906} will be important for the future study of this and related Kitaev spin-liquid candidate materials.

\section*{Acknowledgement}
This work was supported by the Materials Sciences and Engineering Division, Basic Energy Sciences, Office of Science, US Department of Energy. We gratefully acknowledge the computing resources provided on Bebop and Improv, high-performance computing clusters operated by the Laboratory Computing Resource Center at Argonne National Laboratory.

\bibliography{references.bib}

\begin{thebibliography}{51}%
\makeatletter
\providecommand \@ifxundefined [1]{%
 \@ifx{#1\undefined}
}%
\providecommand \@ifnum [1]{%
 \ifnum #1\expandafter \@firstoftwo
 \else \expandafter \@secondoftwo
 \fi
}%
\providecommand \@ifx [1]{%
 \ifx #1\expandafter \@firstoftwo
 \else \expandafter \@secondoftwo
 \fi
}%
\providecommand \natexlab [1]{#1}%
\providecommand \enquote  [1]{``#1''}%
\providecommand \bibnamefont  [1]{#1}%
\providecommand \bibfnamefont [1]{#1}%
\providecommand \citenamefont [1]{#1}%
\providecommand \href@noop [0]{\@secondoftwo}%
\providecommand \href [0]{\begingroup \@sanitize@url \@href}%
\providecommand \@href[1]{\@@startlink{#1}\@@href}%
\providecommand \@@href[1]{\endgroup#1\@@endlink}%
\providecommand \@sanitize@url [0]{\catcode `\\12\catcode `\$12\catcode `\&12\catcode `\#12\catcode `\^12\catcode `\_12\catcode `\%12\relax}%
\providecommand \@@startlink[1]{}%
\providecommand \@@endlink[0]{}%
\providecommand \url  [0]{\begingroup\@sanitize@url \@url }%
\providecommand \@url [1]{\endgroup\@href {#1}{\urlprefix }}%
\providecommand \urlprefix  [0]{URL }%
\providecommand \Eprint [0]{\href }%
\providecommand \doibase [0]{http://dx.doi.org/}%
\providecommand \selectlanguage [0]{\@gobble}%
\providecommand \bibinfo  [0]{\@secondoftwo}%
\providecommand \bibfield  [0]{\@secondoftwo}%
\providecommand \translation [1]{[#1]}%
\providecommand \BibitemOpen [0]{}%
\providecommand \bibitemStop [0]{}%
\providecommand \bibitemNoStop [0]{.\EOS\space}%
\providecommand \EOS [0]{\spacefactor3000\relax}%
\providecommand \BibitemShut  [1]{\csname bibitem#1\endcsname}%
\let\auto@bib@innerbib\@empty
\bibitem [{\citenamefont {Takagi}\ \emph {et~al.}(2019)\citenamefont {Takagi}, \citenamefont {Takayama}, \citenamefont {Jackeli}, \citenamefont {Khaliullin},\ and\ \citenamefont {Nagler}}]{Takagi2019KitaevQS}%
  \BibitemOpen
  \bibfield  {author} {\bibinfo {author} {\bibfnamefont {H.}~\bibnamefont {Takagi}}, \bibinfo {author} {\bibfnamefont {T.}~\bibnamefont {Takayama}}, \bibinfo {author} {\bibfnamefont {G.}~\bibnamefont {Jackeli}}, \bibinfo {author} {\bibfnamefont {G.}~\bibnamefont {Khaliullin}}, \ and\ \bibinfo {author} {\bibfnamefont {S.~E.}\ \bibnamefont {Nagler}},\ }\href {\doibase 10.1038/s42254-019-0038-2} {\bibfield  {journal} {\bibinfo  {journal} {Nature Reviews Physics}\ }\textbf {\bibinfo {volume} {1}},\ \bibinfo {pages} {264} (\bibinfo {year} {2019})}\BibitemShut {NoStop}%
\bibitem [{\citenamefont {Chaloupka}\ \emph {et~al.}(2010)\citenamefont {Chaloupka}, \citenamefont {Jackeli},\ and\ \citenamefont {Khaliullin}}]{PhysRevLett.105.027204}%
  \BibitemOpen
  \bibfield  {author} {\bibinfo {author} {\bibfnamefont {J.~c.~v.}\ \bibnamefont {Chaloupka}}, \bibinfo {author} {\bibfnamefont {G.}~\bibnamefont {Jackeli}}, \ and\ \bibinfo {author} {\bibfnamefont {G.}~\bibnamefont {Khaliullin}},\ }\href {\doibase 10.1103/PhysRevLett.105.027204} {\bibfield  {journal} {\bibinfo  {journal} {Phys. Rev. Lett.}\ }\textbf {\bibinfo {volume} {105}},\ \bibinfo {pages} {027204} (\bibinfo {year} {2010})}\BibitemShut {NoStop}%
\bibitem [{\citenamefont {Liu}\ and\ \citenamefont {Khaliullin}(2018)}]{QSL_Condition_1}%
  \BibitemOpen
  \bibfield  {author} {\bibinfo {author} {\bibfnamefont {H.}~\bibnamefont {Liu}}\ and\ \bibinfo {author} {\bibfnamefont {G.}~\bibnamefont {Khaliullin}},\ }\href {\doibase 10.1103/PhysRevB.97.014407} {\bibfield  {journal} {\bibinfo  {journal} {Phys. Rev. B}\ }\textbf {\bibinfo {volume} {97}},\ \bibinfo {pages} {014407} (\bibinfo {year} {2018})}\BibitemShut {NoStop}%
\bibitem [{\citenamefont {Sano}\ \emph {et~al.}(2018)\citenamefont {Sano}, \citenamefont {Kato},\ and\ \citenamefont {Motome}}]{QSL_Condition_2}%
  \BibitemOpen
  \bibfield  {author} {\bibinfo {author} {\bibfnamefont {R.}~\bibnamefont {Sano}}, \bibinfo {author} {\bibfnamefont {Y.}~\bibnamefont {Kato}}, \ and\ \bibinfo {author} {\bibfnamefont {Y.}~\bibnamefont {Motome}},\ }\href {\doibase 10.1103/PhysRevB.97.014408} {\bibfield  {journal} {\bibinfo  {journal} {Phys. Rev. B}\ }\textbf {\bibinfo {volume} {97}},\ \bibinfo {pages} {014408} (\bibinfo {year} {2018})}\BibitemShut {NoStop}%
\bibitem [{\citenamefont {Liu}\ \emph {et~al.}(2020)\citenamefont {Liu}, \citenamefont {Chaloupka},\ and\ \citenamefont {Khaliullin}}]{PhysRevLett.125.047201}%
  \BibitemOpen
  \bibfield  {author} {\bibinfo {author} {\bibfnamefont {H.}~\bibnamefont {Liu}}, \bibinfo {author} {\bibfnamefont {J.~c.~v.}\ \bibnamefont {Chaloupka}}, \ and\ \bibinfo {author} {\bibfnamefont {G.}~\bibnamefont {Khaliullin}},\ }\href {\doibase 10.1103/PhysRevLett.125.047201} {\bibfield  {journal} {\bibinfo  {journal} {Phys. Rev. Lett.}\ }\textbf {\bibinfo {volume} {125}},\ \bibinfo {pages} {047201} (\bibinfo {year} {2020})}\BibitemShut {NoStop}%
\bibitem [{\citenamefont {Viciu}\ \emph {et~al.}(2007)\citenamefont {Viciu}, \citenamefont {Huang}, \citenamefont {Morosan}, \citenamefont {Zandbergen}, \citenamefont {Greenbaum}, \citenamefont {McQueen},\ and\ \citenamefont {Cava}}]{VICIU}%
  \BibitemOpen
  \bibfield  {author} {\bibinfo {author} {\bibfnamefont {L.}~\bibnamefont {Viciu}}, \bibinfo {author} {\bibfnamefont {Q.}~\bibnamefont {Huang}}, \bibinfo {author} {\bibfnamefont {E.}~\bibnamefont {Morosan}}, \bibinfo {author} {\bibfnamefont {H.}~\bibnamefont {Zandbergen}}, \bibinfo {author} {\bibfnamefont {N.}~\bibnamefont {Greenbaum}}, \bibinfo {author} {\bibfnamefont {T.}~\bibnamefont {McQueen}}, \ and\ \bibinfo {author} {\bibfnamefont {R.}~\bibnamefont {Cava}},\ }\href {\doibase https://doi.org/10.1016/j.jssc.2007.01.002} {\bibfield  {journal} {\bibinfo  {journal} {Journal of Solid State Chemistry}\ }\textbf {\bibinfo {volume} {180}},\ \bibinfo {pages} {1060} (\bibinfo {year} {2007})}\BibitemShut {NoStop}%
\bibitem [{\citenamefont {Wong}\ \emph {et~al.}(2016)\citenamefont {Wong}, \citenamefont {Avdeev},\ and\ \citenamefont {Ling}}]{WONG}%
  \BibitemOpen
  \bibfield  {author} {\bibinfo {author} {\bibfnamefont {C.}~\bibnamefont {Wong}}, \bibinfo {author} {\bibfnamefont {M.}~\bibnamefont {Avdeev}}, \ and\ \bibinfo {author} {\bibfnamefont {C.~D.}\ \bibnamefont {Ling}},\ }\href {\doibase https://doi.org/10.1016/j.jssc.2016.07.032} {\bibfield  {journal} {\bibinfo  {journal} {Journal of Solid State Chemistry}\ }\textbf {\bibinfo {volume} {243}},\ \bibinfo {pages} {18} (\bibinfo {year} {2016})}\BibitemShut {NoStop}%
\bibitem [{\citenamefont {Politaev}\ \emph {et~al.}(2010)\citenamefont {Politaev}, \citenamefont {Nalbandyan}, \citenamefont {Petrenko}, \citenamefont {Shukaev}, \citenamefont {Volotchaev},\ and\ \citenamefont {Medvedev}}]{POLITAEV}%
  \BibitemOpen
  \bibfield  {author} {\bibinfo {author} {\bibfnamefont {V.}~\bibnamefont {Politaev}}, \bibinfo {author} {\bibfnamefont {V.}~\bibnamefont {Nalbandyan}}, \bibinfo {author} {\bibfnamefont {A.}~\bibnamefont {Petrenko}}, \bibinfo {author} {\bibfnamefont {I.}~\bibnamefont {Shukaev}}, \bibinfo {author} {\bibfnamefont {V.}~\bibnamefont {Volotchaev}}, \ and\ \bibinfo {author} {\bibfnamefont {B.}~\bibnamefont {Medvedev}},\ }\href {\doibase https://doi.org/10.1016/j.jssc.2009.12.002} {\bibfield  {journal} {\bibinfo  {journal} {Journal of Solid State Chemistry}\ }\textbf {\bibinfo {volume} {183}},\ \bibinfo {pages} {684} (\bibinfo {year} {2010})}\BibitemShut {NoStop}%
\bibitem [{\citenamefont {Yan}\ \emph {et~al.}(2019)\citenamefont {Yan}, \citenamefont {Okamoto}, \citenamefont {Wu}, \citenamefont {Zheng}, \citenamefont {Zhou}, \citenamefont {Cao},\ and\ \citenamefont {McGuire}}]{PhysRevMaterials.3.074405}%
  \BibitemOpen
  \bibfield  {author} {\bibinfo {author} {\bibfnamefont {J.-Q.}\ \bibnamefont {Yan}}, \bibinfo {author} {\bibfnamefont {S.}~\bibnamefont {Okamoto}}, \bibinfo {author} {\bibfnamefont {Y.}~\bibnamefont {Wu}}, \bibinfo {author} {\bibfnamefont {Q.}~\bibnamefont {Zheng}}, \bibinfo {author} {\bibfnamefont {H.~D.}\ \bibnamefont {Zhou}}, \bibinfo {author} {\bibfnamefont {H.~B.}\ \bibnamefont {Cao}}, \ and\ \bibinfo {author} {\bibfnamefont {M.~A.}\ \bibnamefont {McGuire}},\ }\href {\doibase 10.1103/PhysRevMaterials.3.074405} {\bibfield  {journal} {\bibinfo  {journal} {Phys. Rev. Mater.}\ }\textbf {\bibinfo {volume} {3}},\ \bibinfo {pages} {074405} (\bibinfo {year} {2019})}\BibitemShut {NoStop}%
\bibitem [{\citenamefont {Liu}\ and\ \citenamefont {Kee}(2023)}]{PhysRevB.107.054420}%
  \BibitemOpen
  \bibfield  {author} {\bibinfo {author} {\bibfnamefont {X.}~\bibnamefont {Liu}}\ and\ \bibinfo {author} {\bibfnamefont {H.-Y.}\ \bibnamefont {Kee}},\ }\href {\doibase 10.1103/PhysRevB.107.054420} {\bibfield  {journal} {\bibinfo  {journal} {Phys. Rev. B}\ }\textbf {\bibinfo {volume} {107}},\ \bibinfo {pages} {054420} (\bibinfo {year} {2023})}\BibitemShut {NoStop}%
\bibitem [{\citenamefont {Gretarsson}\ \emph {et~al.}(2024)\citenamefont {Gretarsson}, \citenamefont {Fujihara}, \citenamefont {Sato}, \citenamefont {Gotou}, \citenamefont {Imai}, \citenamefont {Ohgushi}, \citenamefont {Keimer},\ and\ \citenamefont {Suzuki}}]{PhysRevB.109.L180413}%
  \BibitemOpen
  \bibfield  {author} {\bibinfo {author} {\bibfnamefont {H.}~\bibnamefont {Gretarsson}}, \bibinfo {author} {\bibfnamefont {H.}~\bibnamefont {Fujihara}}, \bibinfo {author} {\bibfnamefont {F.}~\bibnamefont {Sato}}, \bibinfo {author} {\bibfnamefont {H.}~\bibnamefont {Gotou}}, \bibinfo {author} {\bibfnamefont {Y.}~\bibnamefont {Imai}}, \bibinfo {author} {\bibfnamefont {K.}~\bibnamefont {Ohgushi}}, \bibinfo {author} {\bibfnamefont {B.}~\bibnamefont {Keimer}}, \ and\ \bibinfo {author} {\bibfnamefont {H.}~\bibnamefont {Suzuki}},\ }\href {\doibase 10.1103/PhysRevB.109.L180413} {\bibfield  {journal} {\bibinfo  {journal} {Phys. Rev. B}\ }\textbf {\bibinfo {volume} {109}},\ \bibinfo {pages} {L180413} (\bibinfo {year} {2024})}\BibitemShut {NoStop}%
\bibitem [{\citenamefont {Kim}\ \emph {et~al.}(2024)\citenamefont {Kim}, \citenamefont {Park}, \citenamefont {Samanta}, \citenamefont {Choi}, \citenamefont {Kang}, \citenamefont {Seo}, \citenamefont {Ji}, \citenamefont {Noh}, \citenamefont {Cho}, \citenamefont {Yoo}, \citenamefont {Ok}, \citenamefont {Kim},\ and\ \citenamefont {Sohn}}]{XAS_TCF}%
  \BibitemOpen
  \bibfield  {author} {\bibinfo {author} {\bibfnamefont {G.-H.}\ \bibnamefont {Kim}}, \bibinfo {author} {\bibfnamefont {M.}~\bibnamefont {Park}}, \bibinfo {author} {\bibfnamefont {S.}~\bibnamefont {Samanta}}, \bibinfo {author} {\bibfnamefont {U.}~\bibnamefont {Choi}}, \bibinfo {author} {\bibfnamefont {B.}~\bibnamefont {Kang}}, \bibinfo {author} {\bibfnamefont {U.}~\bibnamefont {Seo}}, \bibinfo {author} {\bibfnamefont {G.}~\bibnamefont {Ji}}, \bibinfo {author} {\bibfnamefont {S.}~\bibnamefont {Noh}}, \bibinfo {author} {\bibfnamefont {D.-Y.}\ \bibnamefont {Cho}}, \bibinfo {author} {\bibfnamefont {J.-W.}\ \bibnamefont {Yoo}}, \bibinfo {author} {\bibfnamefont {J.~M.}\ \bibnamefont {Ok}}, \bibinfo {author} {\bibfnamefont {H.-S.}\ \bibnamefont {Kim}}, \ and\ \bibinfo {author} {\bibfnamefont {C.}~\bibnamefont {Sohn}},\ }\href {\doibase 10.1126/sciadv.adn8694} {\bibfield  {journal} {\bibinfo  {journal} {Science Advances}\ }\textbf {\bibinfo {volume} {10}},\ \bibinfo {pages} {eadn8694} (\bibinfo {year}
  {2024})}\BibitemShut {NoStop}%
\bibitem [{\citenamefont {van Veenendaal}\ \emph {et~al.}(2023)\citenamefont {van Veenendaal}, \citenamefont {Poldi}, \citenamefont {Veiga}, \citenamefont {Bencok}, \citenamefont {Fabbris}, \citenamefont {Tartaglia}, \citenamefont {McChesney}, \citenamefont {Freeland}, \citenamefont {Hemley}, \citenamefont {Zheng}, \citenamefont {Mitchell}, \citenamefont {Yan},\ and\ \citenamefont {Haskel}}]{PhysRevB.107.214443}%
  \BibitemOpen
  \bibfield  {author} {\bibinfo {author} {\bibfnamefont {M.}~\bibnamefont {van Veenendaal}}, \bibinfo {author} {\bibfnamefont {E.~H.~T.}\ \bibnamefont {Poldi}}, \bibinfo {author} {\bibfnamefont {L.~S.~I.}\ \bibnamefont {Veiga}}, \bibinfo {author} {\bibfnamefont {P.}~\bibnamefont {Bencok}}, \bibinfo {author} {\bibfnamefont {G.}~\bibnamefont {Fabbris}}, \bibinfo {author} {\bibfnamefont {R.}~\bibnamefont {Tartaglia}}, \bibinfo {author} {\bibfnamefont {J.~L.}\ \bibnamefont {McChesney}}, \bibinfo {author} {\bibfnamefont {J.~W.}\ \bibnamefont {Freeland}}, \bibinfo {author} {\bibfnamefont {R.~J.}\ \bibnamefont {Hemley}}, \bibinfo {author} {\bibfnamefont {H.}~\bibnamefont {Zheng}}, \bibinfo {author} {\bibfnamefont {J.~F.}\ \bibnamefont {Mitchell}}, \bibinfo {author} {\bibfnamefont {J.-Q.}\ \bibnamefont {Yan}}, \ and\ \bibinfo {author} {\bibfnamefont {D.}~\bibnamefont {Haskel}},\ }\href {\doibase 10.1103/PhysRevB.107.214443} {\bibfield  {journal} {\bibinfo  {journal} {Phys. Rev. B}\ }\textbf {\bibinfo {volume}
  {107}},\ \bibinfo {pages} {214443} (\bibinfo {year} {2023})}\BibitemShut {NoStop}%
\bibitem [{\citenamefont {Kim}\ \emph {et~al.}(2021{\natexlab{a}})\citenamefont {Kim}, \citenamefont {Jeong}, \citenamefont {Lin}, \citenamefont {Park}, \citenamefont {Masuda}, \citenamefont {Asai}, \citenamefont {Itoh}, \citenamefont {Kim}, \citenamefont {Zhou}, \citenamefont {Ma},\ and\ \citenamefont {Park}}]{INS_TCF}%
  \BibitemOpen
  \bibfield  {author} {\bibinfo {author} {\bibfnamefont {C.}~\bibnamefont {Kim}}, \bibinfo {author} {\bibfnamefont {J.}~\bibnamefont {Jeong}}, \bibinfo {author} {\bibfnamefont {G.}~\bibnamefont {Lin}}, \bibinfo {author} {\bibfnamefont {P.}~\bibnamefont {Park}}, \bibinfo {author} {\bibfnamefont {T.}~\bibnamefont {Masuda}}, \bibinfo {author} {\bibfnamefont {S.}~\bibnamefont {Asai}}, \bibinfo {author} {\bibfnamefont {S.}~\bibnamefont {Itoh}}, \bibinfo {author} {\bibfnamefont {H.-S.}\ \bibnamefont {Kim}}, \bibinfo {author} {\bibfnamefont {H.}~\bibnamefont {Zhou}}, \bibinfo {author} {\bibfnamefont {J.}~\bibnamefont {Ma}}, \ and\ \bibinfo {author} {\bibfnamefont {J.-G.}\ \bibnamefont {Park}},\ }\href {\doibase 10.1088/1361-648X/ac2644} {\bibfield  {journal} {\bibinfo  {journal} {Journal of Physics: Condensed Matter}\ }\textbf {\bibinfo {volume} {34}},\ \bibinfo {pages} {045802} (\bibinfo {year} {2021}{\natexlab{a}})}\BibitemShut {NoStop}%
\bibitem [{\citenamefont {Ferrenti}\ \emph {et~al.}(2023)\citenamefont {Ferrenti}, \citenamefont {Siegler}, \citenamefont {Ghosh}, \citenamefont {Zhang}, \citenamefont {Kintop}, \citenamefont {Vivanco}, \citenamefont {Lygouras}, \citenamefont {Halloran}, \citenamefont {Klemenz}, \citenamefont {Broholm}, \citenamefont {Drichko},\ and\ \citenamefont {McQueen}}]{PhysRevB.108.064433}%
  \BibitemOpen
  \bibfield  {author} {\bibinfo {author} {\bibfnamefont {A.~M.}\ \bibnamefont {Ferrenti}}, \bibinfo {author} {\bibfnamefont {M.~A.}\ \bibnamefont {Siegler}}, \bibinfo {author} {\bibfnamefont {S.}~\bibnamefont {Ghosh}}, \bibinfo {author} {\bibfnamefont {X.}~\bibnamefont {Zhang}}, \bibinfo {author} {\bibfnamefont {N.}~\bibnamefont {Kintop}}, \bibinfo {author} {\bibfnamefont {H.~K.}\ \bibnamefont {Vivanco}}, \bibinfo {author} {\bibfnamefont {C.}~\bibnamefont {Lygouras}}, \bibinfo {author} {\bibfnamefont {T.}~\bibnamefont {Halloran}}, \bibinfo {author} {\bibfnamefont {S.}~\bibnamefont {Klemenz}}, \bibinfo {author} {\bibfnamefont {C.}~\bibnamefont {Broholm}}, \bibinfo {author} {\bibfnamefont {N.}~\bibnamefont {Drichko}}, \ and\ \bibinfo {author} {\bibfnamefont {T.~M.}\ \bibnamefont {McQueen}},\ }\href {\doibase 10.1103/PhysRevB.108.064433} {\bibfield  {journal} {\bibinfo  {journal} {Phys. Rev. B}\ }\textbf {\bibinfo {volume} {108}},\ \bibinfo {pages} {064433} (\bibinfo {year} {2023})}\BibitemShut {NoStop}%
\bibitem [{\citenamefont {Mou}\ \emph {et~al.}(2024)\citenamefont {Mou}, \citenamefont {Zhang}, \citenamefont {Xiang}, \citenamefont {Xu}, \citenamefont {Zhong}, \citenamefont {Cava}, \citenamefont {Zhou}, \citenamefont {Jiang}, \citenamefont {Smirnov}, \citenamefont {Drichko},\ and\ \citenamefont {Winter}}]{10Dq}%
  \BibitemOpen
  \bibfield  {author} {\bibinfo {author} {\bibfnamefont {B.~S.}\ \bibnamefont {Mou}}, \bibinfo {author} {\bibfnamefont {X.}~\bibnamefont {Zhang}}, \bibinfo {author} {\bibfnamefont {L.}~\bibnamefont {Xiang}}, \bibinfo {author} {\bibfnamefont {Y.}~\bibnamefont {Xu}}, \bibinfo {author} {\bibfnamefont {R.}~\bibnamefont {Zhong}}, \bibinfo {author} {\bibfnamefont {R.~J.}\ \bibnamefont {Cava}}, \bibinfo {author} {\bibfnamefont {H.}~\bibnamefont {Zhou}}, \bibinfo {author} {\bibfnamefont {Z.}~\bibnamefont {Jiang}}, \bibinfo {author} {\bibfnamefont {D.}~\bibnamefont {Smirnov}}, \bibinfo {author} {\bibfnamefont {N.}~\bibnamefont {Drichko}}, \ and\ \bibinfo {author} {\bibfnamefont {S.~M.}\ \bibnamefont {Winter}},\ }\href {\doibase 10.1103/PhysRevMaterials.8.084408} {\bibfield  {journal} {\bibinfo  {journal} {Phys. Rev. Mater.}\ }\textbf {\bibinfo {volume} {8}},\ \bibinfo {pages} {084408} (\bibinfo {year} {2024})}\BibitemShut {NoStop}%
\bibitem [{\citenamefont {Nair}\ \emph {et~al.}(2018)\citenamefont {Nair}, \citenamefont {Brown}, \citenamefont {Coldren}, \citenamefont {Hester}, \citenamefont {Gelfand}, \citenamefont {Podlesnyak}, \citenamefont {Huang},\ and\ \citenamefont {Ross}}]{PhysRevB.97.134409}%
  \BibitemOpen
  \bibfield  {author} {\bibinfo {author} {\bibfnamefont {H.~S.}\ \bibnamefont {Nair}}, \bibinfo {author} {\bibfnamefont {J.~M.}\ \bibnamefont {Brown}}, \bibinfo {author} {\bibfnamefont {E.}~\bibnamefont {Coldren}}, \bibinfo {author} {\bibfnamefont {G.}~\bibnamefont {Hester}}, \bibinfo {author} {\bibfnamefont {M.~P.}\ \bibnamefont {Gelfand}}, \bibinfo {author} {\bibfnamefont {A.}~\bibnamefont {Podlesnyak}}, \bibinfo {author} {\bibfnamefont {Q.}~\bibnamefont {Huang}}, \ and\ \bibinfo {author} {\bibfnamefont {K.~A.}\ \bibnamefont {Ross}},\ }\href {\doibase 10.1103/PhysRevB.97.134409} {\bibfield  {journal} {\bibinfo  {journal} {Phys. Rev. B}\ }\textbf {\bibinfo {volume} {97}},\ \bibinfo {pages} {134409} (\bibinfo {year} {2018})}\BibitemShut {NoStop}%
\bibitem [{\citenamefont {Wang}\ \emph {et~al.}(2023)\citenamefont {Wang}, \citenamefont {Sharma}, \citenamefont {Becker}, \citenamefont {Bohat\'y},\ and\ \citenamefont {Lorenz}}]{PhysRevMaterials.7.024402}%
  \BibitemOpen
  \bibfield  {author} {\bibinfo {author} {\bibfnamefont {X.}~\bibnamefont {Wang}}, \bibinfo {author} {\bibfnamefont {R.}~\bibnamefont {Sharma}}, \bibinfo {author} {\bibfnamefont {P.}~\bibnamefont {Becker}}, \bibinfo {author} {\bibfnamefont {L.}~\bibnamefont {Bohat\'y}}, \ and\ \bibinfo {author} {\bibfnamefont {T.}~\bibnamefont {Lorenz}},\ }\href {\doibase 10.1103/PhysRevMaterials.7.024402} {\bibfield  {journal} {\bibinfo  {journal} {Phys. Rev. Mater.}\ }\textbf {\bibinfo {volume} {7}},\ \bibinfo {pages} {024402} (\bibinfo {year} {2023})}\BibitemShut {NoStop}%
\bibitem [{\citenamefont {Samanta}\ \emph {et~al.}(2024{\natexlab{a}})\citenamefont {Samanta}, \citenamefont {Hong},\ and\ \citenamefont {Kim}}]{nano14010009}%
  \BibitemOpen
  \bibfield  {author} {\bibinfo {author} {\bibfnamefont {S.}~\bibnamefont {Samanta}}, \bibinfo {author} {\bibfnamefont {D.}~\bibnamefont {Hong}}, \ and\ \bibinfo {author} {\bibfnamefont {H.-S.}\ \bibnamefont {Kim}},\ }\href {\doibase 10.3390/nano14010009} {\bibfield  {journal} {\bibinfo  {journal} {Nanomaterials}\ }\textbf {\bibinfo {volume} {14}} (\bibinfo {year} {2024}{\natexlab{a}}),\ 10.3390/nano14010009}\BibitemShut {NoStop}%
\bibitem [{\citenamefont {Mitsumoto}\ \emph {et~al.}(2022)\citenamefont {Mitsumoto}, \citenamefont {Hotta},\ and\ \citenamefont {Yoshino}}]{JT-ice}%
  \BibitemOpen
  \bibfield  {author} {\bibinfo {author} {\bibfnamefont {K.}~\bibnamefont {Mitsumoto}}, \bibinfo {author} {\bibfnamefont {C.}~\bibnamefont {Hotta}}, \ and\ \bibinfo {author} {\bibfnamefont {H.}~\bibnamefont {Yoshino}},\ }\href {\doibase 10.1103/PhysRevResearch.4.033157} {\bibfield  {journal} {\bibinfo  {journal} {Phys. Rev. Res.}\ }\textbf {\bibinfo {volume} {4}},\ \bibinfo {pages} {033157} (\bibinfo {year} {2022})}\BibitemShut {NoStop}%
\bibitem [{\citenamefont {Liechtenstein}\ \emph {et~al.}(1995)\citenamefont {Liechtenstein}, \citenamefont {Anisimov},\ and\ \citenamefont {Zaanen}}]{LDA+U}%
  \BibitemOpen
  \bibfield  {author} {\bibinfo {author} {\bibfnamefont {A.~I.}\ \bibnamefont {Liechtenstein}}, \bibinfo {author} {\bibfnamefont {V.~I.}\ \bibnamefont {Anisimov}}, \ and\ \bibinfo {author} {\bibfnamefont {J.}~\bibnamefont {Zaanen}},\ }\href {\doibase 10.1103/PhysRevB.52.R5467} {\bibfield  {journal} {\bibinfo  {journal} {Phys. Rev. B}\ }\textbf {\bibinfo {volume} {52}},\ \bibinfo {pages} {R5467} (\bibinfo {year} {1995})}\BibitemShut {NoStop}%
\bibitem [{\citenamefont {Bl\"ochl}(1994)}]{PAW}%
  \BibitemOpen
  \bibfield  {author} {\bibinfo {author} {\bibfnamefont {P.~E.}\ \bibnamefont {Bl\"ochl}},\ }\href {\doibase 10.1103/PhysRevB.50.17953} {\bibfield  {journal} {\bibinfo  {journal} {Phys. Rev. B}\ }\textbf {\bibinfo {volume} {50}},\ \bibinfo {pages} {17953} (\bibinfo {year} {1994})}\BibitemShut {NoStop}%
\bibitem [{\citenamefont {Kresse}\ and\ \citenamefont {Furthm\"uller}(1996)}]{vasp1}%
  \BibitemOpen
  \bibfield  {author} {\bibinfo {author} {\bibfnamefont {G.}~\bibnamefont {Kresse}}\ and\ \bibinfo {author} {\bibfnamefont {J.}~\bibnamefont {Furthm\"uller}},\ }\href {\doibase 10.1103/PhysRevB.54.11169} {\bibfield  {journal} {\bibinfo  {journal} {Phys. Rev. B}\ }\textbf {\bibinfo {volume} {54}},\ \bibinfo {pages} {11169} (\bibinfo {year} {1996})}\BibitemShut {NoStop}%
\bibitem [{\citenamefont {Kresse}\ and\ \citenamefont {Joubert}(1999)}]{vasp2}%
  \BibitemOpen
  \bibfield  {author} {\bibinfo {author} {\bibfnamefont {G.}~\bibnamefont {Kresse}}\ and\ \bibinfo {author} {\bibfnamefont {D.}~\bibnamefont {Joubert}},\ }\href {\doibase 10.1103/PhysRevB.59.1758} {\bibfield  {journal} {\bibinfo  {journal} {Phys. Rev. B}\ }\textbf {\bibinfo {volume} {59}},\ \bibinfo {pages} {1758} (\bibinfo {year} {1999})}\BibitemShut {NoStop}%
\bibitem [{\citenamefont {Perdew}\ \emph {et~al.}(1997)\citenamefont {Perdew}, \citenamefont {Burke},\ and\ \citenamefont {Ernzerhof}}]{PBE}%
  \BibitemOpen
  \bibfield  {author} {\bibinfo {author} {\bibfnamefont {J.~P.}\ \bibnamefont {Perdew}}, \bibinfo {author} {\bibfnamefont {K.}~\bibnamefont {Burke}}, \ and\ \bibinfo {author} {\bibfnamefont {M.}~\bibnamefont {Ernzerhof}},\ }\href {\doibase 10.1103/PhysRevLett.78.1396} {\bibfield  {journal} {\bibinfo  {journal} {Phys. Rev. Lett.}\ }\textbf {\bibinfo {volume} {78}},\ \bibinfo {pages} {1396} (\bibinfo {year} {1997})}\BibitemShut {NoStop}%
\bibitem [{\citenamefont {Singh}\ \emph {et~al.}(2021)\citenamefont {Singh}, \citenamefont {Herath}, \citenamefont {Wah}, \citenamefont {Liao}, \citenamefont {Romero},\ and\ \citenamefont {Park}}]{SINGH2021107778}%
  \BibitemOpen
  \bibfield  {author} {\bibinfo {author} {\bibfnamefont {V.}~\bibnamefont {Singh}}, \bibinfo {author} {\bibfnamefont {U.}~\bibnamefont {Herath}}, \bibinfo {author} {\bibfnamefont {B.}~\bibnamefont {Wah}}, \bibinfo {author} {\bibfnamefont {X.}~\bibnamefont {Liao}}, \bibinfo {author} {\bibfnamefont {A.~H.}\ \bibnamefont {Romero}}, \ and\ \bibinfo {author} {\bibfnamefont {H.}~\bibnamefont {Park}},\ }\href {\doibase https://doi.org/10.1016/j.cpc.2020.107778} {\bibfield  {journal} {\bibinfo  {journal} {Computer Physics Communications}\ }\textbf {\bibinfo {volume} {261}},\ \bibinfo {pages} {107778} (\bibinfo {year} {2021})}\BibitemShut {NoStop}%
\bibitem [{\citenamefont {Mostofi}\ \emph {et~al.}(2008)\citenamefont {Mostofi}, \citenamefont {Yates}, \citenamefont {Lee}, \citenamefont {Souza}, \citenamefont {Vanderbilt},\ and\ \citenamefont {Marzari}}]{Wannier90}%
  \BibitemOpen
  \bibfield  {author} {\bibinfo {author} {\bibfnamefont {A.~A.}\ \bibnamefont {Mostofi}}, \bibinfo {author} {\bibfnamefont {J.~R.}\ \bibnamefont {Yates}}, \bibinfo {author} {\bibfnamefont {Y.-S.}\ \bibnamefont {Lee}}, \bibinfo {author} {\bibfnamefont {I.}~\bibnamefont {Souza}}, \bibinfo {author} {\bibfnamefont {D.}~\bibnamefont {Vanderbilt}}, \ and\ \bibinfo {author} {\bibfnamefont {N.}~\bibnamefont {Marzari}},\ }\href {\doibase 10.1016/j.cpc.2007.11.016} {\bibfield  {journal} {\bibinfo  {journal} {Computer Physics Communications}\ }\textbf {\bibinfo {volume} {178}},\ \bibinfo {pages} {685–699} (\bibinfo {year} {2008})}\BibitemShut {NoStop}%
\bibitem [{\citenamefont {Lee}\ \emph {et~al.}(2023)\citenamefont {Lee}, \citenamefont {Park},\ and\ \citenamefont {Ngo}}]{LeePRB2023}%
  \BibitemOpen
  \bibfield  {author} {\bibinfo {author} {\bibfnamefont {A.~T.}\ \bibnamefont {Lee}}, \bibinfo {author} {\bibfnamefont {H.}~\bibnamefont {Park}}, \ and\ \bibinfo {author} {\bibfnamefont {A.~T.}\ \bibnamefont {Ngo}},\ }\href {\doibase 10.1103/PhysRevB.108.205146} {\bibfield  {journal} {\bibinfo  {journal} {Phys. Rev. B}\ }\textbf {\bibinfo {volume} {108}},\ \bibinfo {pages} {205146} (\bibinfo {year} {2023})}\BibitemShut {NoStop}%
\bibitem [{\citenamefont {Haule}(2007)}]{CTQMC4}%
  \BibitemOpen
  \bibfield  {author} {\bibinfo {author} {\bibfnamefont {K.}~\bibnamefont {Haule}},\ }\href {\doibase 10.1103/PhysRevB.75.155113} {\bibfield  {journal} {\bibinfo  {journal} {Phys. Rev. B}\ }\textbf {\bibinfo {volume} {75}},\ \bibinfo {pages} {155113} (\bibinfo {year} {2007})}\BibitemShut {NoStop}%
\bibitem [{\citenamefont {Kim}\ \emph {et~al.}(2019)\citenamefont {Kim}, \citenamefont {Haule},\ and\ \citenamefont {Vanderbilt}}]{Single-Site-DMFT}%
  \BibitemOpen
  \bibfield  {author} {\bibinfo {author} {\bibfnamefont {H.-S.}\ \bibnamefont {Kim}}, \bibinfo {author} {\bibfnamefont {K.}~\bibnamefont {Haule}}, \ and\ \bibinfo {author} {\bibfnamefont {D.}~\bibnamefont {Vanderbilt}},\ }\href {\doibase 10.1103/PhysRevLett.123.236401} {\bibfield  {journal} {\bibinfo  {journal} {Phys. Rev. Lett.}\ }\textbf {\bibinfo {volume} {123}},\ \bibinfo {pages} {236401} (\bibinfo {year} {2019})}\BibitemShut {NoStop}%
\bibitem [{\citenamefont {Samanta}\ \emph {et~al.}(2024{\natexlab{b}})\citenamefont {Samanta}, \citenamefont {Cossu},\ and\ \citenamefont {Kim}}]{Samanta2024-pi}%
  \BibitemOpen
  \bibfield  {author} {\bibinfo {author} {\bibfnamefont {S.}~\bibnamefont {Samanta}}, \bibinfo {author} {\bibfnamefont {F.}~\bibnamefont {Cossu}}, \ and\ \bibinfo {author} {\bibfnamefont {H.-S.}\ \bibnamefont {Kim}},\ }\href@noop {} {\bibfield  {journal} {\bibinfo  {journal} {Npj Quantum Mater.}\ }\textbf {\bibinfo {volume} {9}} (\bibinfo {year} {2024}{\natexlab{b}})}\BibitemShut {NoStop}%
\bibitem [{\citenamefont {Georgescu}\ \emph {et~al.}(2022)\citenamefont {Georgescu}, \citenamefont {Millis},\ and\ \citenamefont {Rondinelli}}]{PhysRevB.105.245153}%
  \BibitemOpen
  \bibfield  {author} {\bibinfo {author} {\bibfnamefont {A.~B.}\ \bibnamefont {Georgescu}}, \bibinfo {author} {\bibfnamefont {A.~J.}\ \bibnamefont {Millis}}, \ and\ \bibinfo {author} {\bibfnamefont {J.~M.}\ \bibnamefont {Rondinelli}},\ }\href {\doibase 10.1103/PhysRevB.105.245153} {\bibfield  {journal} {\bibinfo  {journal} {Phys. Rev. B}\ }\textbf {\bibinfo {volume} {105}},\ \bibinfo {pages} {245153} (\bibinfo {year} {2022})}\BibitemShut {NoStop}%
\bibitem [{\citenamefont {Rodolakis}\ \emph {et~al.}(2011)\citenamefont {Rodolakis}, \citenamefont {Rueff}, \citenamefont {Sikora}, \citenamefont {Alliot}, \citenamefont {Iti\'e}, \citenamefont {Baudelet}, \citenamefont {Ravy}, \citenamefont {Wzietek}, \citenamefont {Hansmann}, \citenamefont {Toschi}, \citenamefont {Haverkort}, \citenamefont {Sangiovanni}, \citenamefont {Held}, \citenamefont {Metcalf},\ and\ \citenamefont {Marsi}}]{Orbital_Shape}%
  \BibitemOpen
  \bibfield  {author} {\bibinfo {author} {\bibfnamefont {F.}~\bibnamefont {Rodolakis}}, \bibinfo {author} {\bibfnamefont {J.-P.}\ \bibnamefont {Rueff}}, \bibinfo {author} {\bibfnamefont {M.}~\bibnamefont {Sikora}}, \bibinfo {author} {\bibfnamefont {I.}~\bibnamefont {Alliot}}, \bibinfo {author} {\bibfnamefont {J.-P.}\ \bibnamefont {Iti\'e}}, \bibinfo {author} {\bibfnamefont {F.}~\bibnamefont {Baudelet}}, \bibinfo {author} {\bibfnamefont {S.}~\bibnamefont {Ravy}}, \bibinfo {author} {\bibfnamefont {P.}~\bibnamefont {Wzietek}}, \bibinfo {author} {\bibfnamefont {P.}~\bibnamefont {Hansmann}}, \bibinfo {author} {\bibfnamefont {A.}~\bibnamefont {Toschi}}, \bibinfo {author} {\bibfnamefont {M.~W.}\ \bibnamefont {Haverkort}}, \bibinfo {author} {\bibfnamefont {G.}~\bibnamefont {Sangiovanni}}, \bibinfo {author} {\bibfnamefont {K.}~\bibnamefont {Held}}, \bibinfo {author} {\bibfnamefont {P.}~\bibnamefont {Metcalf}}, \ and\ \bibinfo {author} {\bibfnamefont {M.}~\bibnamefont {Marsi}},\ }\href {\doibase
  10.1103/PhysRevB.84.245113} {\bibfield  {journal} {\bibinfo  {journal} {Phys. Rev. B}\ }\textbf {\bibinfo {volume} {84}},\ \bibinfo {pages} {245113} (\bibinfo {year} {2011})}\BibitemShut {NoStop}%
\bibitem [{\citenamefont {Winter}(2022)}]{Winter_2022}%
  \BibitemOpen
  \bibfield  {author} {\bibinfo {author} {\bibfnamefont {S.~M.}\ \bibnamefont {Winter}},\ }\href {\doibase 10.1088/2515-7639/ac94f8} {\bibfield  {journal} {\bibinfo  {journal} {Journal of Physics: Materials}\ }\textbf {\bibinfo {volume} {5}},\ \bibinfo {pages} {045003} (\bibinfo {year} {2022})}\BibitemShut {NoStop}%
\bibitem [{\citenamefont {Nguyen}\ \emph {et~al.}(2021)\citenamefont {Nguyen}, \citenamefont {Yamauchi}, \citenamefont {Oguchi}, \citenamefont {Amoroso},\ and\ \citenamefont {Picozzi}}]{PhysRevB.104.014414}%
  \BibitemOpen
  \bibfield  {author} {\bibinfo {author} {\bibfnamefont {T.~P.~T.}\ \bibnamefont {Nguyen}}, \bibinfo {author} {\bibfnamefont {K.}~\bibnamefont {Yamauchi}}, \bibinfo {author} {\bibfnamefont {T.}~\bibnamefont {Oguchi}}, \bibinfo {author} {\bibfnamefont {D.}~\bibnamefont {Amoroso}}, \ and\ \bibinfo {author} {\bibfnamefont {S.}~\bibnamefont {Picozzi}},\ }\href {\doibase 10.1103/PhysRevB.104.014414} {\bibfield  {journal} {\bibinfo  {journal} {Phys. Rev. B}\ }\textbf {\bibinfo {volume} {104}},\ \bibinfo {pages} {014414} (\bibinfo {year} {2021})}\BibitemShut {NoStop}%
\bibitem [{\citenamefont {Katsnelson}\ and\ \citenamefont {Lichtenstein}(2000)}]{PhysRevB.61.8906}%
  \BibitemOpen
  \bibfield  {author} {\bibinfo {author} {\bibfnamefont {M.~I.}\ \bibnamefont {Katsnelson}}\ and\ \bibinfo {author} {\bibfnamefont {A.~I.}\ \bibnamefont {Lichtenstein}},\ }\href {\doibase 10.1103/PhysRevB.61.8906} {\bibfield  {journal} {\bibinfo  {journal} {Phys. Rev. B}\ }\textbf {\bibinfo {volume} {61}},\ \bibinfo {pages} {8906} (\bibinfo {year} {2000})}\BibitemShut {NoStop}%
\bibitem [{\citenamefont {Marzari}\ \emph {et~al.}(2012)\citenamefont {Marzari}, \citenamefont {Mostofi}, \citenamefont {Yates}, \citenamefont {Souza},\ and\ \citenamefont {Vanderbilt}}]{Marzari}%
  \BibitemOpen
  \bibfield  {author} {\bibinfo {author} {\bibfnamefont {N.}~\bibnamefont {Marzari}}, \bibinfo {author} {\bibfnamefont {A.~A.}\ \bibnamefont {Mostofi}}, \bibinfo {author} {\bibfnamefont {J.~R.}\ \bibnamefont {Yates}}, \bibinfo {author} {\bibfnamefont {I.}~\bibnamefont {Souza}}, \ and\ \bibinfo {author} {\bibfnamefont {D.}~\bibnamefont {Vanderbilt}},\ }\href {\doibase 10.1103/revmodphys.84.1419} {\bibfield  {journal} {\bibinfo  {journal} {Reviews of Modern Physics}\ }\textbf {\bibinfo {volume} {84}},\ \bibinfo {pages} {1419–1475} (\bibinfo {year} {2012})}\BibitemShut {NoStop}%
\bibitem [{\citenamefont {Mostofi}\ \emph {et~al.}(2014)\citenamefont {Mostofi}, \citenamefont {Yates}, \citenamefont {Pizzi}, \citenamefont {Lee}, \citenamefont {Souza}, \citenamefont {Vanderbilt},\ and\ \citenamefont {Marzari}}]{MOSTOFI20142309}%
  \BibitemOpen
  \bibfield  {author} {\bibinfo {author} {\bibfnamefont {A.~A.}\ \bibnamefont {Mostofi}}, \bibinfo {author} {\bibfnamefont {J.~R.}\ \bibnamefont {Yates}}, \bibinfo {author} {\bibfnamefont {G.}~\bibnamefont {Pizzi}}, \bibinfo {author} {\bibfnamefont {Y.-S.}\ \bibnamefont {Lee}}, \bibinfo {author} {\bibfnamefont {I.}~\bibnamefont {Souza}}, \bibinfo {author} {\bibfnamefont {D.}~\bibnamefont {Vanderbilt}}, \ and\ \bibinfo {author} {\bibfnamefont {N.}~\bibnamefont {Marzari}},\ }\href {\doibase https://doi.org/10.1016/j.cpc.2014.05.003} {\bibfield  {journal} {\bibinfo  {journal} {Computer Physics Communications}\ }\textbf {\bibinfo {volume} {185}},\ \bibinfo {pages} {2309} (\bibinfo {year} {2014})}\BibitemShut {NoStop}%
\bibitem [{\citenamefont {Kim}\ \emph {et~al.}(2021{\natexlab{b}})\citenamefont {Kim}, \citenamefont {Kim},\ and\ \citenamefont {Park}}]{SL_Cond1}%
  \BibitemOpen
  \bibfield  {author} {\bibinfo {author} {\bibfnamefont {C.}~\bibnamefont {Kim}}, \bibinfo {author} {\bibfnamefont {H.-S.}\ \bibnamefont {Kim}}, \ and\ \bibinfo {author} {\bibfnamefont {J.-G.}\ \bibnamefont {Park}},\ }\href {\doibase 10.1088/1361-648X/ac2d5d} {\bibfield  {journal} {\bibinfo  {journal} {Journal of Physics: Condensed Matter}\ }\textbf {\bibinfo {volume} {34}},\ \bibinfo {pages} {023001} (\bibinfo {year} {2021}{\natexlab{b}})}\BibitemShut {NoStop}%
\bibitem [{\citenamefont {Held}(2007)}]{DMFT1}%
  \BibitemOpen
  \bibfield  {author} {\bibinfo {author} {\bibfnamefont {K.}~\bibnamefont {Held}},\ }\href {\doibase 10.1080/00018730701619647} {\bibfield  {journal} {\bibinfo  {journal} {Advances in Physics}\ }\textbf {\bibinfo {volume} {56}},\ \bibinfo {pages} {829} (\bibinfo {year} {2007})}\BibitemShut {NoStop}%
\bibitem [{\citenamefont {Anderson}(1961)}]{AIM}%
  \BibitemOpen
  \bibfield  {author} {\bibinfo {author} {\bibfnamefont {P.~W.}\ \bibnamefont {Anderson}},\ }\href {\doibase 10.1103/PhysRev.124.41} {\bibfield  {journal} {\bibinfo  {journal} {Phys. Rev.}\ }\textbf {\bibinfo {volume} {124}},\ \bibinfo {pages} {41} (\bibinfo {year} {1961})}\BibitemShut {NoStop}%
\bibitem [{\citenamefont {Kotliar}\ and\ \citenamefont {Vollhardt}(2004)}]{10.1063/1.1712502}%
  \BibitemOpen
  \bibfield  {author} {\bibinfo {author} {\bibfnamefont {G.}~\bibnamefont {Kotliar}}\ and\ \bibinfo {author} {\bibfnamefont {D.}~\bibnamefont {Vollhardt}},\ }\href {\doibase 10.1063/1.1712502} {\bibfield  {journal} {\bibinfo  {journal} {Physics Today}\ }\textbf {\bibinfo {volume} {57}},\ \bibinfo {pages} {53} (\bibinfo {year} {2004})}\BibitemShut {NoStop}%
\bibitem [{\citenamefont {IV}\ and\ \citenamefont {Ferrero}(2022)}]{ctQMC_issue1}%
  \BibitemOpen
  \bibfield  {author} {\bibinfo {author} {\bibfnamefont {F.~i. c.~v.}\ \bibnamefont {IV}}\ and\ \bibinfo {author} {\bibfnamefont {M.}~\bibnamefont {Ferrero}},\ }\href {\doibase 10.1103/PhysRevB.105.125104} {\bibfield  {journal} {\bibinfo  {journal} {Phys. Rev. B}\ }\textbf {\bibinfo {volume} {105}},\ \bibinfo {pages} {125104} (\bibinfo {year} {2022})}\BibitemShut {NoStop}%
\bibitem [{\citenamefont {Li}\ and\ \citenamefont {Yao}(2019)}]{ctQMC_issue2}%
  \BibitemOpen
  \bibfield  {author} {\bibinfo {author} {\bibfnamefont {Z.-X.}\ \bibnamefont {Li}}\ and\ \bibinfo {author} {\bibfnamefont {H.}~\bibnamefont {Yao}},\ }\href {\doibase 10.1146/annurev-conmatphys-033117-054307} {\bibfield  {journal} {\bibinfo  {journal} {Ann. Rev. Condensed Matter Phys.}\ }\textbf {\bibinfo {volume} {10}},\ \bibinfo {pages} {337} (\bibinfo {year} {2019})},\ \Eprint {http://arxiv.org/abs/1805.08219} {arXiv:1805.08219 [cond-mat.str-el]} \BibitemShut {NoStop}%
\bibitem [{\citenamefont {Sheridan}\ \emph {et~al.}(2019)\citenamefont {Sheridan}, \citenamefont {Weber}, \citenamefont {Plekhanov},\ and\ \citenamefont {Rhodes}}]{CTQMC1}%
  \BibitemOpen
  \bibfield  {author} {\bibinfo {author} {\bibfnamefont {E.}~\bibnamefont {Sheridan}}, \bibinfo {author} {\bibfnamefont {C.}~\bibnamefont {Weber}}, \bibinfo {author} {\bibfnamefont {E.}~\bibnamefont {Plekhanov}}, \ and\ \bibinfo {author} {\bibfnamefont {C.}~\bibnamefont {Rhodes}},\ }\href {\doibase 10.1103/PhysRevB.99.205156} {\bibfield  {journal} {\bibinfo  {journal} {Phys. Rev. B}\ }\textbf {\bibinfo {volume} {99}},\ \bibinfo {pages} {205156} (\bibinfo {year} {2019})}\BibitemShut {NoStop}%
\bibitem [{\citenamefont {Gull}\ \emph {et~al.}(2011)\citenamefont {Gull}, \citenamefont {Millis}, \citenamefont {Lichtenstein}, \citenamefont {Rubtsov}, \citenamefont {Troyer},\ and\ \citenamefont {Werner}}]{CTQMC2}%
  \BibitemOpen
  \bibfield  {author} {\bibinfo {author} {\bibfnamefont {E.}~\bibnamefont {Gull}}, \bibinfo {author} {\bibfnamefont {A.~J.}\ \bibnamefont {Millis}}, \bibinfo {author} {\bibfnamefont {A.~I.}\ \bibnamefont {Lichtenstein}}, \bibinfo {author} {\bibfnamefont {A.~N.}\ \bibnamefont {Rubtsov}}, \bibinfo {author} {\bibfnamefont {M.}~\bibnamefont {Troyer}}, \ and\ \bibinfo {author} {\bibfnamefont {P.}~\bibnamefont {Werner}},\ }\href {\doibase 10.1103/RevModPhys.83.349} {\bibfield  {journal} {\bibinfo  {journal} {Rev. Mod. Phys.}\ }\textbf {\bibinfo {volume} {83}},\ \bibinfo {pages} {349} (\bibinfo {year} {2011})}\BibitemShut {NoStop}%
\bibitem [{\citenamefont {Werner}\ and\ \citenamefont {Millis}(2006)}]{CTQMC3}%
  \BibitemOpen
  \bibfield  {author} {\bibinfo {author} {\bibfnamefont {P.}~\bibnamefont {Werner}}\ and\ \bibinfo {author} {\bibfnamefont {A.~J.}\ \bibnamefont {Millis}},\ }\href {\doibase 10.1103/PhysRevB.74.155107} {\bibfield  {journal} {\bibinfo  {journal} {Phys. Rev. B}\ }\textbf {\bibinfo {volume} {74}},\ \bibinfo {pages} {155107} (\bibinfo {year} {2006})}\BibitemShut {NoStop}%
\bibitem [{\citenamefont {Jarrell}\ and\ \citenamefont {Gubernatis}(1996)}]{Maximum_Entropy}%
  \BibitemOpen
  \bibfield  {author} {\bibinfo {author} {\bibfnamefont {M.}~\bibnamefont {Jarrell}}\ and\ \bibinfo {author} {\bibfnamefont {J.}~\bibnamefont {Gubernatis}},\ }\href {\doibase https://doi.org/10.1016/0370-1573(95)00074-7} {\bibfield  {journal} {\bibinfo  {journal} {Physics Reports}\ }\textbf {\bibinfo {volume} {269}},\ \bibinfo {pages} {133} (\bibinfo {year} {1996})}\BibitemShut {NoStop}%
\bibitem [{\citenamefont {Kang}\ \emph {et~al.}(2023)\citenamefont {Kang}, \citenamefont {Park}, \citenamefont {Song}, \citenamefont {Noh}, \citenamefont {Choe}, \citenamefont {Kong}, \citenamefont {Kim}, \citenamefont {Seo}, \citenamefont {Ko}, \citenamefont {Yi}, \citenamefont {Yoo}, \citenamefont {Park}, \citenamefont {Ok},\ and\ \citenamefont {Sohn}}]{PhysRevB.107.075103}%
  \BibitemOpen
  \bibfield  {author} {\bibinfo {author} {\bibfnamefont {B.}~\bibnamefont {Kang}}, \bibinfo {author} {\bibfnamefont {M.}~\bibnamefont {Park}}, \bibinfo {author} {\bibfnamefont {S.}~\bibnamefont {Song}}, \bibinfo {author} {\bibfnamefont {S.}~\bibnamefont {Noh}}, \bibinfo {author} {\bibfnamefont {D.}~\bibnamefont {Choe}}, \bibinfo {author} {\bibfnamefont {M.}~\bibnamefont {Kong}}, \bibinfo {author} {\bibfnamefont {M.}~\bibnamefont {Kim}}, \bibinfo {author} {\bibfnamefont {C.}~\bibnamefont {Seo}}, \bibinfo {author} {\bibfnamefont {E.~K.}\ \bibnamefont {Ko}}, \bibinfo {author} {\bibfnamefont {G.}~\bibnamefont {Yi}}, \bibinfo {author} {\bibfnamefont {J.-W.}\ \bibnamefont {Yoo}}, \bibinfo {author} {\bibfnamefont {S.}~\bibnamefont {Park}}, \bibinfo {author} {\bibfnamefont {J.~M.}\ \bibnamefont {Ok}}, \ and\ \bibinfo {author} {\bibfnamefont {C.}~\bibnamefont {Sohn}},\ }\href {\doibase 10.1103/PhysRevB.107.075103} {\bibfield  {journal} {\bibinfo  {journal} {Phys. Rev. B}\ }\textbf {\bibinfo {volume} {107}},\ \bibinfo
  {pages} {075103} (\bibinfo {year} {2023})}\BibitemShut {NoStop}%
\bibitem [{\citenamefont {Takayama}\ \emph {et~al.}(2021)\citenamefont {Takayama}, \citenamefont {Chaloupka}, \citenamefont {Smerald}, \citenamefont {Khaliullin},\ and\ \citenamefont {Takagi}}]{Explanation}%
  \BibitemOpen
  \bibfield  {author} {\bibinfo {author} {\bibfnamefont {T.}~\bibnamefont {Takayama}}, \bibinfo {author} {\bibfnamefont {J.}~\bibnamefont {Chaloupka}}, \bibinfo {author} {\bibfnamefont {A.}~\bibnamefont {Smerald}}, \bibinfo {author} {\bibfnamefont {G.}~\bibnamefont {Khaliullin}}, \ and\ \bibinfo {author} {\bibfnamefont {H.}~\bibnamefont {Takagi}},\ }\href {\doibase 10.7566/JPSJ.90.062001} {\bibfield  {journal} {\bibinfo  {journal} {Journal of the Physical Society of Japan}\ }\textbf {\bibinfo {volume} {90}},\ \bibinfo {pages} {062001} (\bibinfo {year} {2021})},\ \Eprint {http://arxiv.org/abs/https://doi.org/10.7566/JPSJ.90.062001} {https://doi.org/10.7566/JPSJ.90.062001} \BibitemShut {NoStop}%
\bibitem [{\citenamefont {Foyevtsova}\ \emph {et~al.}(2013)\citenamefont {Foyevtsova}, \citenamefont {Jeschke}, \citenamefont {Mazin}, \citenamefont {Khomskii},\ and\ \citenamefont {Valent\'{\i}}}]{JpUp}%
  \BibitemOpen
  \bibfield  {author} {\bibinfo {author} {\bibfnamefont {K.}~\bibnamefont {Foyevtsova}}, \bibinfo {author} {\bibfnamefont {H.~O.}\ \bibnamefont {Jeschke}}, \bibinfo {author} {\bibfnamefont {I.~I.}\ \bibnamefont {Mazin}}, \bibinfo {author} {\bibfnamefont {D.~I.}\ \bibnamefont {Khomskii}}, \ and\ \bibinfo {author} {\bibfnamefont {R.}~\bibnamefont {Valent\'{\i}}},\ }\href {\doibase 10.1103/PhysRevB.88.035107} {\bibfield  {journal} {\bibinfo  {journal} {Phys. Rev. B}\ }\textbf {\bibinfo {volume} {88}},\ \bibinfo {pages} {035107} (\bibinfo {year} {2013})}\BibitemShut {NoStop}%
\end{thebibliography}%

\appendix

\section{Magnetic Ground State of Na$_2$Co$_3$SbO$_6$}
\label{Sec:Energetics}

\begin{figure}[htb!]
  \centering
  \includegraphics[width=0.35\textwidth]{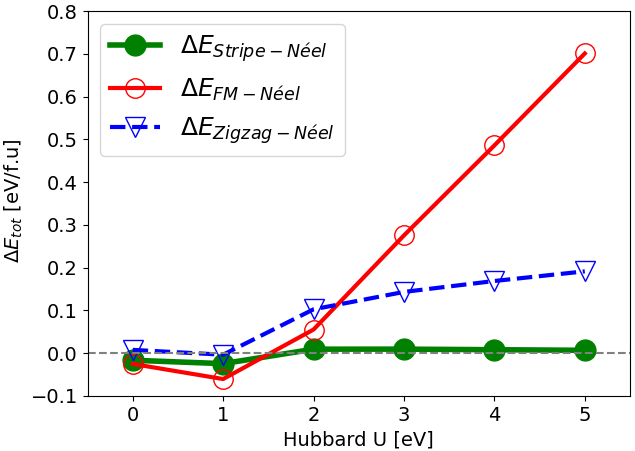}
\caption{Relative energy per formula unit of FM, zigzag AFM, stripe AFM structure to N\'{e}el AFM structure of NCSO as a function of the Hubbard $U$ while fixing $J$=0.8 eV within DFT+U calculation.}   
  \label{ground_state}
\end{figure}

Here, we briefly discuss about magnetic ground states of Na$_2$Co$_3$SbO$_6$ using DFT+U while varying $U$ and fixing $J$=0.8 eV. The energy differences between magnetic structures, including FM, zigzag AFM, stripe AFM, and the N\'{e}el AFM show that the N\'{e}el AFM structure is more stable than the other magnetic structures in a strongly correlated regime ($U>2$ eV), while the FM is the most stable structure in a weakly correlated regime (Fig. \ref{ground_state}).

\vspace{-0.8cm}

\section{Wannier Hamiltonian}

Here, we construct the tight-binding Hamiltonian using the maximally localized Wannier functions (MLWFs)~\cite{Marzari}. 
The correlated subspace of Co 3$d$ orbitals can be solved using the DMFT equations.
Trigonal and Jahn-Teller distortions in Na$_3$Co$_2$SbO$_6$ can also be studied using the on-site orbital energies of Co 3$d$ orbitals. 
To construct Wannier functions, we first solve the non-spin-polarized Kohn-Sham equation within DFT using VASP code. Once the Kohn-Sham equation is solved, all eigenfunctions $\psi_{n\bm{k}}(\bm{r})$ and the energy bands are obtained: $E_n(\bm{k})=\bra{\psi_{n\bm{k}}}\hat{H}^{KS}\ket{\psi_{n\bm{k}}}$, where $\hat{H}^{KS}$ is the Kohn-Sham Hamiltonian. We then constructed a manifold of $J$ bands with maximally localized Wannier functions (MLWFs)\cite{Marzari} 
\begin{align}
\ket{\bm{R}n}=\frac{V}{(2\pi)^3}\int d\bm{k}e^{-i\bm{k}\cdot\bm{R}}\sum_{m=1}^{J}\mathcal{U}_{mn}^{(\bm{k})}\ket{\psi_{m\bm{k}}}
\end{align}
using VASP and WANNIER90\cite{MOSTOFI20142309} codes. In this work, the local [001], [010], and [001] axes for constructing Co $d$ orbitals are chosen be aligned close to the local bondings of the perfect octahedron such that the local [111] axis will point toward the $c-$axis. After the Wannier procedure, the real-space Wannier Hamiltonian $\mathcal{H}_{mn}(\bm{R}'-\bm{R})=\bra{\bm{R}'m}\hat{H}^{KS}\ket{\bm{R}n}$ is obtained.

The local Wannier Hamiltonians ($\mathcal{H}_{mn}(\bm{0})$) of the Co $3d$ orbitals of NCSO (within the $spd$-model), BaCo$_2($AsO$_4$)$_2$ with the space group $R\bar{3}$ [No.\:148] (within the $spd$-model), and the CoI$_2$ with the space group $P\bar{3}m1$ [No.\:164] (within the $pd$-model) respectively at zero strain are represented using the cubic harmonic basis
($\{\ket{z^2},\ket{x^2-y^2},\ket{xz},\ket{yz}\},\ket{xy}\}$): 
\begin{align}
\label{NCSO_spd_cubic}
\mathcal{N}^{spd}_{0}=\left[\begin{array}{cc|ccc}
4.756&  0.000&    -0.005& -0.005&  0.009\\
0.000&     4.756& -0.011&  0.011& 0.000\\
\hline
-0.005& -0.011&  4.437&  0.004&  0.004\\
-0.005&  0.011&  0.004&  4.437&  0.004\\
0.009& 0.000&     0.004&  0.004&  4.440
\end{array}
\right]
\\
\label{BCAO_spd_cubic}
\mathcal{B}^{spd}_{0}=\left[\begin{array}{cc|ccc}
2.860&   0.000&    -0.004&  0.007& -0.002\\
 0.000&     2.860&   0.005&  0.002& -0.006\\
 \hline
 -0.004&  0.005&  2.598&  0.014&  0.014\\
 0.007&  0.002&  0.014&  2.597&  0.014\\
 -0.002& -0.006&  0.014&  0.014&  2.598
\end{array}
\right]
\end{align}
\begin{align}
\label{CI_pd_cubic}
\mathcal{C}^{pd}_{0}=\left[\begin{array}{cc|ccc}
4.056& 0.000&    -0.001& -0.001&  0.003\\
 0.000&     4.056& -0.002&  0.002&  0.000\\
 \hline
 -0.001& -0.002&  3.887&  0.007&  0.007\\
 -0.001&  0.002&  0.007&  3.887&  0.007\\
0.003& -0.000&     0.007&  0.007&  3.887
\end{array}
\right]
\end{align}

By using a unitary matrix 
\begin{align}
\label{U_trigonal}
U_{trig}=\left[\begin{array}{c|ccccc}
&\ket{e_g^{\sigma}}& \ket{e_g^{\sigma}}& \ket{a_{1g}}& \ket{e_{g+}^{\pi}}& \ket{e_{g-}^{\pi}}\\
\hline
\bra{z^2} & 1.0 & 0.0 & 0.0 & 0.0 & 0.0\\
\bra{x^2-y^2} & 0.0 & 1.0 & 0.0 & 0.0 & 0.0 \\ 
\bra{xz} & 0.0  & 0.0 & \frac{1}{\sqrt{3}} & -\frac{1}{\sqrt{6}} & \frac{1}{\sqrt{2}}\\
\bra{yz} & 0.0 & 0.0 & \frac{1}{\sqrt{3}} & -\frac{1}{\sqrt{6}} & -\frac{1}{\sqrt{2}} \\
\bra{xy} & 0.0 & 0.0& \frac{1}{\sqrt{3}} & \frac{2}{\sqrt{6}} & 0.0 \\
\end{array}
\right]
\end{align}
that rotates the cubic harmonic basis to the trigonal basis $\{\ket{e_g^{\sigma}},\ket{e_g^{\sigma}},\ket{a_{1g}},\ket{e_{g+}^{\pi}}\},\ket{e_{g-}^{\pi}}\}$, the matrices (\ref{NCSO_spd_cubic}),(\ref{BCAO_spd_cubic}), and (\ref{CI_pd_cubic}) become  
\begin{align}
\label{NCSO_spd_trigonal}
\big[\mathcal{N}^{spd}_{0}\big]_{tri}&=\left[\begin{array}{cc|ccc}
4.756&  0.000&     0.001&  0.011& -0.007\\
0.000&   4.756& -0.002& -0.01&  -0.016\\
\hline
0.001& -0.002&  4.446&  0.002&  0.000   \\
0.011& -0.010&   0.002&  4.434&  0.000   \\
-0.007& -0.016&  0.00&     0.000&     4.433
\end{array}
\right]
\\
\label{BCAO_spd_trigonal}
\big[\mathcal{B}^{spd}_{0}\big]_{tri}&=\left[\begin{array}{cc|ccc}
2.860&   0.000&    0.000&    -0.003& -0.008\\
 0.000&     2.860&   0.000&    -0.008&  0.002\\
 \hline
 0.000&     0.000&     2.626&  0.000& 0.000\\
 -0.003& -0.008&  0.000&     2.584&  0.000   \\
 -0.008&  0.002&  0.000&     0.000&     2.583
\end{array}
\right]
\\
\label{CI_pd_trigonal}
\big[\mathcal{C}^{spd}_{0}\big]_{tri}&=\left[\begin{array}{cc|ccc}
4.056&  0.000&     0.000&     0.004&  0.000\\
0.000&  4.056&     0.000&     0.000&    -0.003\\
\hline
0.000&  0.000&     3.901&     0.000&     0.000\\
0.004&  0.000&     0.000&     3.880&   0.000\\   
0.000& -0.003&     0.000&     0.000&     3.880 
\end{array}
\right]
\end{align}
in the trigonal basis $\{\ket{e_g^{\sigma}},\ket{e_g^{\sigma}},\ket{a_{1g}},\ket{e_{g+}^{\pi}}\},\ket{e_{g-}^{\pi}}\}$.

The $\{\ket{a_{1g}},\ket{e_{g+}^{\pi}}\},\ket{e_{g-}^{\pi}}\}$ blocks of the matrices (\ref{BCAO_spd_trigonal}) and (\ref{CI_pd_trigonal}) have the off-diagonal terms of zeros, implying the trigonal basis is more suitable than the cubic one for BCAO and CoI$_2$. This is consistent since both of the space groups $R\bar{3}$ of BaCo$_2($AsO$_4$)$_2$ and $P\bar{3}m1$ of CoI$_2$ represent for trigonal systems although the $R\bar{3}$ has a lower symmetry than the $P\bar{3}m1$. On the other hand, the $\{\ket{a_{1g}},\ket{e_{g+}^{\pi}}\},\ket{e_{g-}^{\pi}}\}$ block of the matrix (\ref{NCSO_spd_trigonal}) has a small non-zero off-diagonal terms, originated from the small JT compression at zero strain  due to the monoclinic space group $C2/m$ of NCSO.

To see the effects of both trigonal and JT distortions within the space group $C2/m$, we consider a Hamiltonian  
\begin{align}
\label{H}
\mathcal{H}=\left[\begin{array}{cc|ccc}
E_1 & 0.0 & \epsilon_1 & \epsilon_2 & \epsilon_3\\
0.0 & E_1+\delta_1 & \epsilon_4  & \epsilon_5 & \epsilon_6 \\ 
\hline
\epsilon_1  & \epsilon_4 & E_2 & \frac{\Delta_{tri}}{3}+\delta_3 & \frac{\Delta_{tri}}{3}\\
\epsilon_2 & \epsilon_5 & \frac{\Delta_{tri}}{3}+\delta_3 & E_2 & \frac{\Delta_{tri}}{3} \\
\epsilon_3 & \epsilon_6 & \frac{\Delta_{tri}}{3} & \frac{\Delta_{tri}}{3} & E_2+\delta_2 \\
\end{array}
\right]
\end{align}
in the cubic basis $\{\ket{z^2},\ket{x^2-y^2},\ket{xz},\ket{yz}\},\ket{xy}\}$ where $E_1$=$E_{d_{z^2}}$ and $E_2$=$E_{d_{xz}}$=$E_{d_{yz}}$. The parameters $\delta_1$, $\delta_2$, $\delta_3$  represent the energy splitting between $d_{z^2}$ and $d_{x^2-y^2}$ orbitals, the one between $d_{xz}$ and $d_{xy}$ orbitals, and the distortion between (Co-O)$_{\hat{x}}$ and (Co-O)$_{\hat{y}}$ away from 90$^{\degree}$ respectively, due to the JT effect from the monoclinic-cell type.
The parameter $\frac{\Delta_{tri}}{3}$ represents the trigonal distortion.  

By using the unitary matrix (\ref{U_trigonal}), one can obtain the matrix representation of $\mathcal{H}$ in the trigonal basis such that

\begin{widetext}
\begin{align}
\label{Eff. Ham. in trigonal basis}
[\mathcal{H}]_{trig}=U_{trig}^{\dagger}[\mathcal{H}]U_{trig}= 
\left[\begin{array}{cc|ccc}
E_1 & 0.0 & \epsilon_1' & \epsilon_2' & \epsilon_3'\\
0.0 & E_1+\delta_1 & \epsilon_4'  & \epsilon_5' & \epsilon_6' \\ 
\hline
\epsilon_1'  & \epsilon_4' & E_{2}+\frac{2}{3}\Delta_{tri}+\frac{\delta_2}{3}+\frac{2}{3}\delta_3 & \frac{\sqrt{2}}{3}(\delta_2-\delta_3) & 0.0\\
 \epsilon_2' & \epsilon_5' & \frac{\sqrt{2}}{3}(\delta_2-\delta_3) & E_{2}-\frac{1}{3}\Delta_{tri}+\frac{2}{3}\delta_2+\frac{\delta_3}{3} & 0.0 \\
 \epsilon_3' & \epsilon_6' & 0.0 & 0.0 & E_2-\frac{1}{3}\Delta_{tri}-\delta_3 \\
\end{array}
\right]
\end{align}
\end{widetext}

Then the trigonal crystal field $\Delta_{TCF}$, defined as the energy splitting between the singlet $a_{1g}$ and the doublet $e_{g}^{\pi}$ in the trigonal basis, is given by 
\begin{align}
\label{TCF formula}
\Delta_{TCF}=E_{a_{1g}}-\frac{E_{e_{g+}'}+E_{e_{g-}'}}{2}=\Delta_{tri}+\delta_{3} 
\end{align}
Basically, the sum of all the upper (lower) off-diagonal terms in the $t_{2g}$ block from the cubic Hamiltonian $\mathcal{H}$ (Eq. (\ref{H})) represents the trigonal crystal field $\Delta_{TCF}$ in the trigonal basis. Also, the effect of JT distortion is given by the energy splitting between $e_{g+}'$ and $e_{g-}'$ such that 
\begin{align}
\label{JT formula}
\delta_{JT}=E_{e_{g+}'}-E_{e_{g-}'}=\frac{2}{3}(\delta_2+2\delta_3)
\end{align}

From Wannier calculations, we also obtained the $d-d$ hopping matrix between the nearest Co neighbors ($Z$-bond) of the zero strain structure within the $d$-model
\begin{align}
\label{T_dd}
\mathcal{T}^{d}_{d-d}=\left[\begin{array}{ccccc}
 -0.055&  0.000& -0.010&  -0.010&     0.117\\
    0.000& -0.006&  0.012&  -0.012&     0.000\\
  -0.010& 0.012&  0.039&  -0.026&      0.032\\
  -0.010& -0.012& -0.026&   0.039&     0.032\\
    0.117&  0.0&  0.0322&   0.032&    -0.162
\end{array}
\right]    
\end{align}

\begin{table*}[htb!]
\caption{Hopping integrals and charge transfer gap $\Delta_{pd}$ of NCSO at zero strain.} 
\begin{tabular}{c c c c c c c c c c c}
\hline
model  & $t_1$[eV]  & $t_2$[eV] & $t_3$[eV] & $t_4$[eV] & $t_5$[eV] & $t_6$[eV] & $t_{pd}^{\sigma}[eV]$& $t_{pd}^{\pi}[eV]$& $\Delta_{pd}[eV]$

  \\ [0.5ex] 
\hline
$d$ & 0.039 & $-0.026$  & $-0.162$
 & $-0.055$ & $-0.006$ & 0.117 & N/A& N/A & N/A
\\

Winter\cite{Winter_2022} & $\sim0.036$ &$\sim -0.03$ &$\sim-0.142$ &$\sim-0.036$ &$\sim-0.036$ &$\sim$ 0.130 &$\sim$ 1
&$\sim$0.5 &4.5
\\
Kim $et$ $al$\cite{SL_Cond1} & N/A& $0.007$&$-0.135$ &N/A &N/A &0.125 &1.179 & -0.589& 4.21
\\
\hline
$spd$ & 0.054 & $-0.040$ 
& $-0.176$ & $-0.019$    & $-0.055$
& $-0.035$ & 1.061 & $-0.534$ & 3.58

\\
\hline
\end{tabular}
\label{hopping integral}
\end{table*} 

\begin{figure}[htb!]
  \centering
  \includegraphics[width=0.35\textwidth]{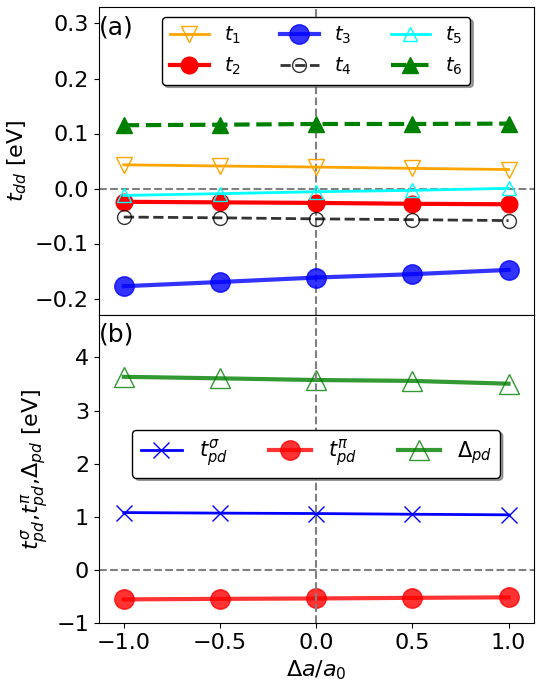}
\caption{Strain effect on (a) the hopping integral between $d$ orbitals and (b) the $p-d$ hopping integral and the charge transfer gap $\Delta_{pd}$.}   
  \label{hopping_charge_transfer}
\end{figure}

By comparing the matrix (\ref{T_dd}) to the hopping matrix of the Co nearest-neighbors ($Z$-bond) for an ideal edge-sharing bond with $90^{\degree}$ metal–ligand–metal bond angles  
\begin{align}
\label{hopping dd}
\mathcal{T}^{dd}_{ij}=\left[\begin{array}{cc ccc}
t_4&  0 &    0 & 0 &  t_6 \\
0 &     t_5& 0 &  0 & 0 \\
0 & 0 &  t_1&  t_2&  0 \\
0 &  0 &  t_2&  t_1&  0\\
t_6 & 0 &     0 &  0 &  t_3
\end{array}
\right]
\end{align}
in the cubic basis of $\{\ket{z^2}$,$\ket{x^2-y^2}$,$\ket{xz},\ket{yz}\},\ket{xy}\}$~\cite{PhysRevB.107.054420,Winter_2022}, 
one can find the hopping integrals for each exchange channel $t_{2g}$-$t_{2g}$ ($t_1$, $t_2$, $t_3$), $e_{g}$-$e_{g}$ ($t_4$, $t_5$), and $t_{2g}$-$e_{g}$ $(t_6)$ at zero strain as shown in Table \ref{hopping integral}. These hopping integrals are comparable with previous $ab$ $initio$ calculations \cite{Winter_2022,SL_Cond1}.

The onsite Hamiltonian for O $2p$ orbitals (within the $spd$-model) is 
\begin{align}
\label{H_2p}
\mathcal{H}^{O-2p}_{0}=\left[\begin{array}{c|ccc}
& \ket{p_z} & \ket{p_x} & \ket{p_y}
\\
\hline
\bra{p_z}& 0.311&   0.117&   0.117\\
\bra{p_x}& 0.117&   1.139&   0.006\\
\bra{p_y}& 0.117&   0.006&   1.139\\
\end{array}
\right]    
\end{align}

By using the matrices (\ref{NCSO_spd_cubic}) and (\ref{H_2p}), one can find the charge transfer gap $\Delta_{pd}=\overline{E}_{t_{2g}}-\overline{E}_{2p}=3.58$ eV at zero strain

Finally, the numerical hopping between $p-d$ orbitals within the $spd$-model is given by 
\begin{align}
\label{T_dd}
\mathcal{T}^{spd}_{p-d}=\left[\begin{array}{c|ccc}
& \ket{p_z} & \ket{p_x} & \ket{p_y}
\\
\hline
\bra{z^2}& 0.095&  -0.537&   0.010\\
\bra{x^2-y^2}& -0.167&   0.908&  -0.030\\
\bra{xz}& -0.516&  -0.194&   0.010\\
\bra{yz}& 0.001&  0.036&   0.053\\
\bra{xy}& -0.009&  -0.021&  -0.559\\
 \end{array}
\right]    
\end{align}

By comparing to the $p$-$d$ hopping matrix for an ideal octahedron~\cite{PhysRevB.107.054420} 
\begin{align}
\label{T_dd}
\mathcal{T}^{0}_{p-d}=\left[\begin{array}{c|ccc}
& \ket{p_z} & \ket{p_x} & \ket{p_y}
\\
\hline
\bra{z^2}& 0.0&  -\frac{1}{2}t_{pd}^\sigma&   0.0\\
\bra{x^2-y^2}& 0.0&   \frac{\sqrt{3}}{2}t_{pd}^\sigma&  0.0\\
\bra{xz}& t_{pd}^\pi&  0.0&   0.0\\
\bra{yz}& 0.0&  0.0&   0.0\\
\bra{xy}& 0.0&  0.0&  t_{pd}^\pi\\
 \end{array}
\right],    
\end{align}
one can obtain $t_{pd}^{\sigma}=\frac{1}{2}\big(2*0.537+\frac{2}{\sqrt{3}}*0.908)=1.061$ eV and $t_{pd}^{\pi}=\frac{1}{2}\big(-0.516--0.559\big)=-0.534$ eV. Figure \ref{hopping_charge_transfer} shows that only the direct $d-d$ hopping integral $t_3$ favoring Heisenberg interaction is sensitive to the strain effect.

\section{Energy Band of the Ambient NCSO Structure}

\begin{figure}[htb!]
    \includegraphics[width=0.35\textwidth]{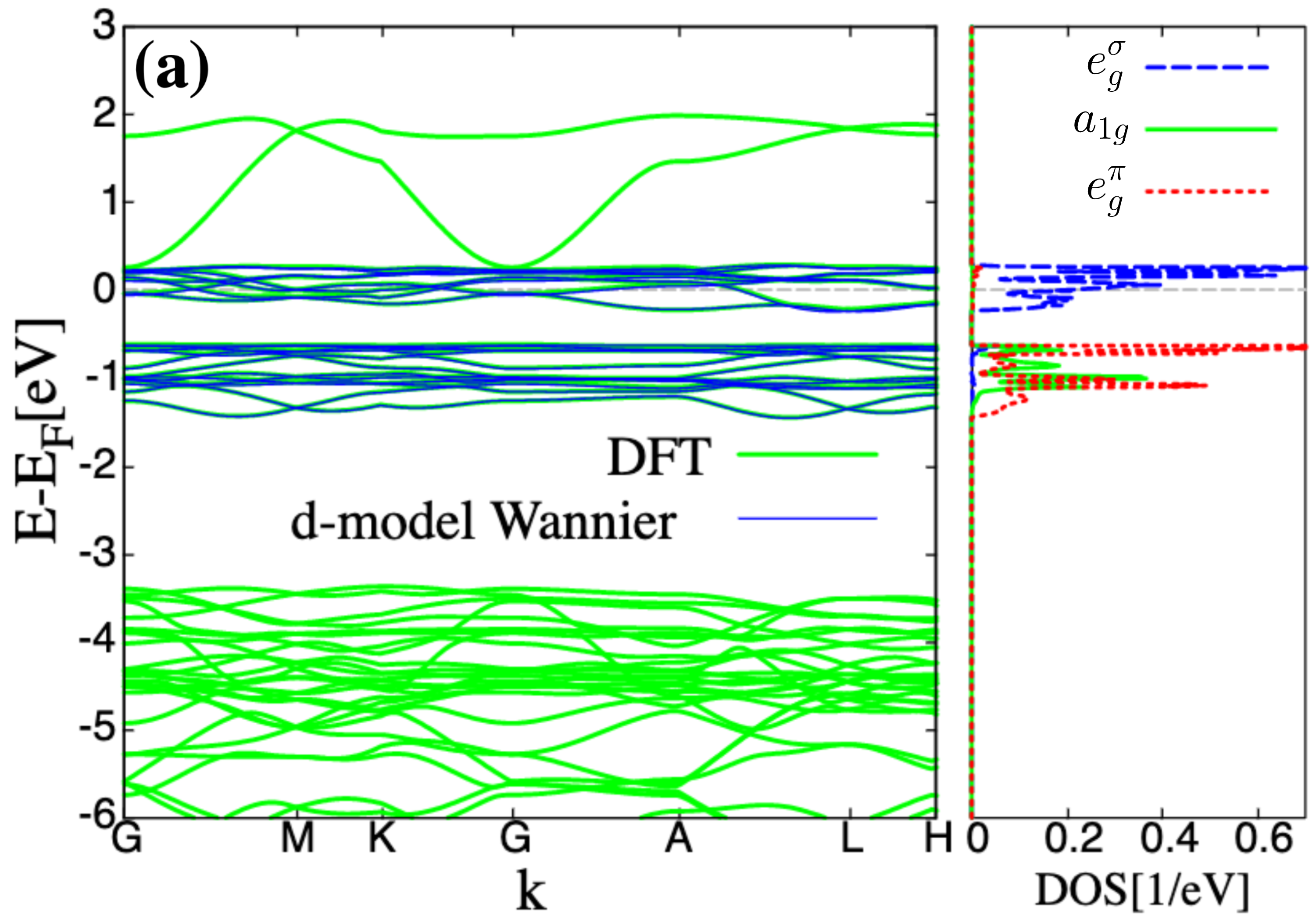}
    \includegraphics[width=0.35\textwidth]{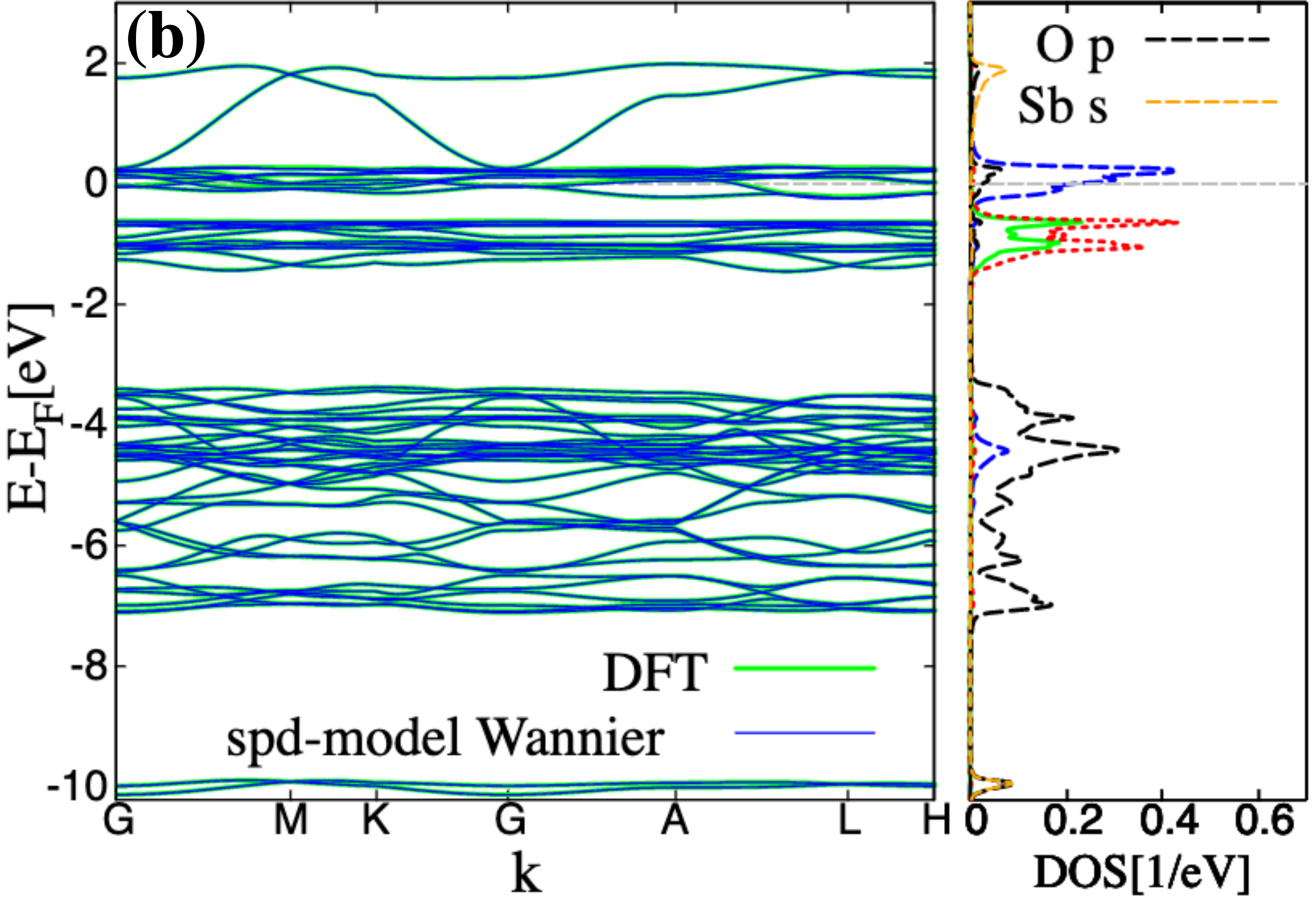} 
    \includegraphics[width=0.35\textwidth]{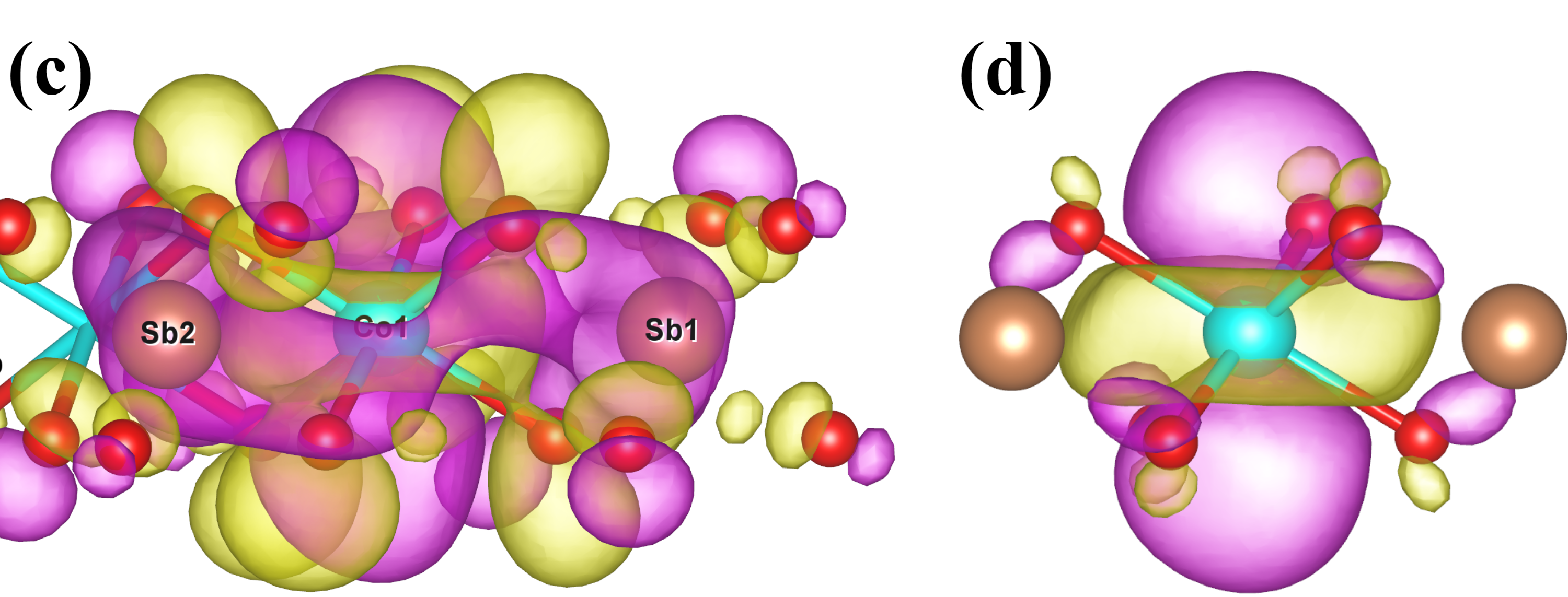} 
\caption{(a,b) Energy bands of NCSO calculated with DFT and Wannier functions. (c) Molecular-like $a_{1g}$ orbital in the trigonal basis under the effects of ligand fields within d-model Wannier functions. (d) Atomic-like $a_{1g}$ orbital in trigonal basis within spd-model Wannier functions}    
   \label{Energy Bands}
\end{figure}

Here, we compare the non-spin-polarized DFT and Wannier energy bands obtained from the fully relaxed NCSO structure. Fig.\:\ref{Energy Bands}(a,b) show that the DFT and Wannier bands fit well for both $d-$ and $spd$-models, indicating that the eigenvalues of the Kohn-Sham equation are exactly the same with those obtained from the $d$/$spd$-model Wannier functions. Using the obtained Wannier functions, the Co $3d$ Hamiltonians can be extracted and analyzed.

\section{DMFT method}
\label{DMFT method}
DMFT maps the lattice many-body problem with the on-site interaction $U$ into a single-site problem with the same interaction $U$ and coupled to self-consistently determined bath $\Delta(i\omega)$~\cite{DMFT1}.
In the limit of the large lattice coordination, the self-energy $\bm{\Sigma} (\bm{k},i\omega)\approx \bm{\Sigma} (i\omega) $, then the local self-energy $\bm{\Sigma}(i\omega)$ can be treated as the single-site impurity problem coupled to electron-bath as in the Anderson impurity model (AIM)\cite{AIM,10.1063/1.1712502}. The hybridization function $\Delta(i\omega)$ describing the ability of an electron to hop in and hop out of an atom is due to the spatial overlap between the correlated orbitals and the conduction band states. 

The DFT+DMFT calculation procedure can be described as follows. First, we diagonalize the $spd$-model Wannier Hamiltonian, such as the matrix (\ref{NCSO_spd_cubic}), in order to obtain the Hamiltonian $H_{D}$ where the Co $d$ submatrix can be diagonal with a correspondent unitary matrix $U_{diag}$ of column eigenvectors. Since the matrix representations of the local Co $3d$ Wannier Hamiltonians in cubic and trigonal bases contain non-zero off-diagonal terms which can lead to the fermion-sign problem in Monte Carlo simulation~\cite{ctQMC_issue1,ctQMC_issue2}, these diagonalized Hamiltonians without off-diagonal terms can be beneficial. All of the final results, including density of states (DOS) (or spectral function), occupancy matrix, and the imaginary part of self-energy $\Sigma(\omega)$ will be rotated from the diagonal basis to the trigonal basis with the unitary matrix $U=U_{diag\rightarrow trig}=U_{diag}^{\dagger}\cdot U_{trig}$. 

Next, the Hamiltonian $H^{D}$ is then used to construct the local lattice Green's function $\bm{G}_{loc}$ within the DMFT self-consistent loop. Within the DMFT self-consistent loop, starting from the local impurity self-energy $\bm{\Sigma}(i\omega)$, one can obtain the matrix of $\bm{G}_{loc}$:
\begin{equation}
\label{lattice_Green_function}
\bm{G}_{loc}(i\omega)=\frac{1}{V_{BZ}}\int_{BZ}d\bm{k}\big[(i\omega+\mu)\mathbb{1}-\mathcal{H}^{D}-(\bm{\Sigma}(i\omega)-V_{DC})\big]^{-1}
\end{equation}
where $\mu$ is the chemical potential, $\mathbb{1}$ is the identity matrix, $V_{DC}$ is the double-counting potential, and $V_{BZ}$ is the volume of the Brillouin zone. We use the fully-localized-limit double-counting potential $V_{DC}$ given by
\begin{align}
V_{DC}=U\bigg(N_d-\frac{1}{2}\bigg)-\frac{J}{2}\big(N_d-1\big)
\end{align}
where $N_d$ is the $d$ occupancy.

Once the lattice Green’s function $\bm{G}_{loc}(i\omega)$ is computed from the Eq.\:\ref{lattice_Green_function}, then the effective non-interacting Green’s function $\bm{\mathcal{G}}^{0}$ of the Anderson model can be found by 
\begin{align}
\label{first}
\big[\bm{\mathcal{G}}_0(i\omega)\big]^{-1}=\big[\bm{G}_{loc}(i\omega)\big]^{-1}+\bm{\Sigma}(i\omega).
\end{align}
This non-interacting Green’s function $\bm{\mathcal{G}}_{0}$ is then used to solve the Anderson impurity model and calculate the interacting impurity Green's function $\bm{G}_{imp}(i\omega$)  using continuous time quantumn Monte Carlo (CTQMC) impurity solver\cite{CTQMC1,CTQMC2,CTQMC3,CTQMC4}. Then, the new self-energy is given by 
\begin{align}
\label{second}
\bm{\Sigma}_{new}(i\omega)=\big[\bm{\mathcal{G}}_0(i\omega)\big]^{-1}-\big[\bm{G}_{imp}(i\omega)\big]^{-1}.
\end{align}
Then, we substitute the mixed new self-energy $\bm{\Sigma}_{new}$ into the (Eq.\:\ref{lattice_Green_function}) so that the algorithm iterates until the convergence criterion is satisfied, e.g.
\begin{align}
\bm{G}_{loc}(i\omega)= \bm{G}_{imp}(i\omega).
\end{align}

One then can obtain the analytically continued Green's function $\bm{G}_{A}$ using the maximum entropy method~\cite{Maximum_Entropy}. Then, the spectral function $\bm{A}(\omega)$ in the trigonal basis, which can be compared to photoemission in experiment, is given by  

\begin{align}
\bm{A}(\omega)=-\frac{1}{\pi}\text{Im}\big[U^{\dagger}\bm{G}_{A}(\omega)U\big]
\end{align}   

Also, the occupancy matrix in the trigonal basis is given by  
\begin{align}
\bm{N}=\int_{\omega}\bm{A}(\omega)f_E(\omega)d\omega
\end{align}
where $f_E(\omega)$ is the Fermi distribution function.

\section{Effect of the different choice of Wannier orbitals}
\label{The effect of hybridization}
\begin{figure}[htb!]
  \centering
  \includegraphics[width=0.35\textwidth]{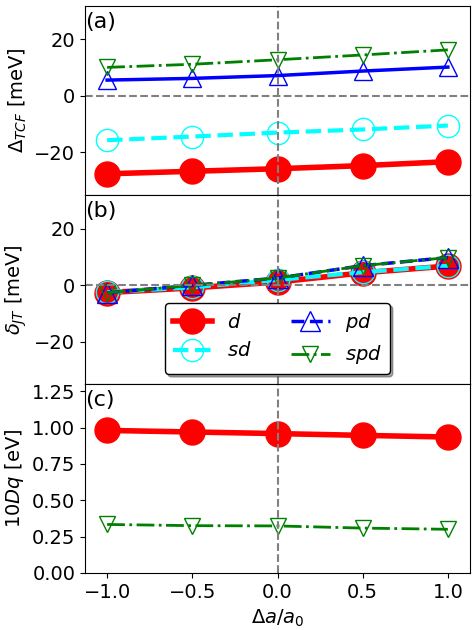}
\caption{(a,b) $\Delta_{TCF}$ and $\delta_{JT}$ of NCSO as a function of strain within $d$-,$sd$-,$pd$-, and $spd$-model Wannier Hamiltonians. (c) Crystal field splitting $10Dq$ between $e_g^{\sigma}$ and $t_{2g}$ orbitals of NCSO within $d$ and $spd$ models.}   
  \label{effect of energy windows}
\end{figure}

To study the effect of different choices of Wannier orbitals on $\Delta_{TCF}$ and $\delta_{JT}$, we plot the $\Delta_{TCF}$ and the energy splitting $\delta_{JT}$ of NCSO obtained using different choices of Co $d$, Sb $s$, and O $p$ orbitals as a function of strain (Fig. \ref{effect of energy windows}(a,b)). We obtained $d$-, $sd$-, $pd$-, and $spd$-model Wannier Hamiltonians with energy windows of [-2.0, 0.8]eV, [-2.0, 2,5]eV, [-8,0, 2.5]eV, and [-10.5, 2.5]eV respectively. Within these models, the slopes of the $\Delta_{TCF}$ as a function of strain are the same to each other but shifted between the models (see Fig.~\ref{effect of energy windows}a).  This implies the hybridization of Wannier obtials and their environments can play a key role in determining the magnitude of $\Delta_{TCF}$. At zero strain, the $\Delta_{TCF}$ within the $d$-model is negative with $\Delta_{TCF}=-25.8$ meV, consistent with the one measured by Kim $et$ $al$ using XAS~\cite{XAS_TCF}. Also, since the geometric lobes of the $e_{g}^{\pi}$ orbitals ($e_{g+}^{\pi},e_{g-}^{\pi}$) are similar to each other, the energy splitting $\delta_{JT}$ between them does not change much across all models.  

The choice of Wannier orbitals also has a similar effect on the crystal field splitting $10Dq$ between $e_g^{\sigma}$ and $t_{2g}$ orbitals (Fig. \ref{effect of energy windows}(c)). At zero strain the $spd$-model $10Dq$ is only about 0.32 eV, while the $d$-model $10Dq$ is about 0.96 eV, consistent with the common crystal field of Co$^{2+}$ in literature \cite{10Dq,PhysRevB.107.214443,PhysRevB.107.075103}. Within the $d$-model, the $e_g^{\sigma}$ orbitals whose lobes directly point to the surrounding O are more strongly repulsed by the ligand fields of these O$^{2-}$ than the $t_{2g}$ orbitals whose lobes points between these O. Thus, the energy levels of the $e_g^{\sigma}$ orbitals are much higher than those of the $t_{2g}$ orbitals. Within the $spd$-model, 
the $e_g^{\sigma}$ orbitals are much localized and $10Dq$ is decreased.

\section{Comparison of
Trigonal Distortions between Na$_{3}$Co$_2$SbO$_6$, BaCo$_2($AsO$_4$)$_2$, and CoI$_2$}

In the main text, we compared the trigonal distortions and the $\Delta_{TCF}$ of NCSO and CoI$_2$. Here, we also show the trigonal distortion and the $\Delta_{TCF}$ of BaCo$_2($AsO$_4$)$_2$ (BCAO). Fig.\:\ref{TCF of NCSO, BCAO, and CoI2}(a) shows the slopes of trigonal distortions ($L_2/L_1$ and $L_3/L_1$) as a function of strain in BCAO is  similar to the ones of NCSO and Co$I_2$. Figure \ref{TCF of NCSO, BCAO, and CoI2}(b) shows that the $\Delta_{TCF}$ of BCAO, whose symmetry is higher (lower) than NCSO (Co$I_2$), is more (less) sensitive to the strain than the one of NCSO (CoI$_2$). These indicates the symmetry (space group) of a compound determines how sensitive the $\Delta_{TCF}$ of the compound is under the effect of strain. Importantly, the positive slopes of the $\Delta_{TCF}$ as a function of strain in NCSO, CoI$_2$, and BCAO yield a unique and crucial effect of strain, in which the energy level of the $a_{1g}$ orbital increases (decreases) relative to the ones of the $e_{g}^{\pi}$ orbitals as these compounds are more tensile (compressive).    
\begin{figure}[htb!]
  \centering
  \includegraphics[width=0.35\textwidth]{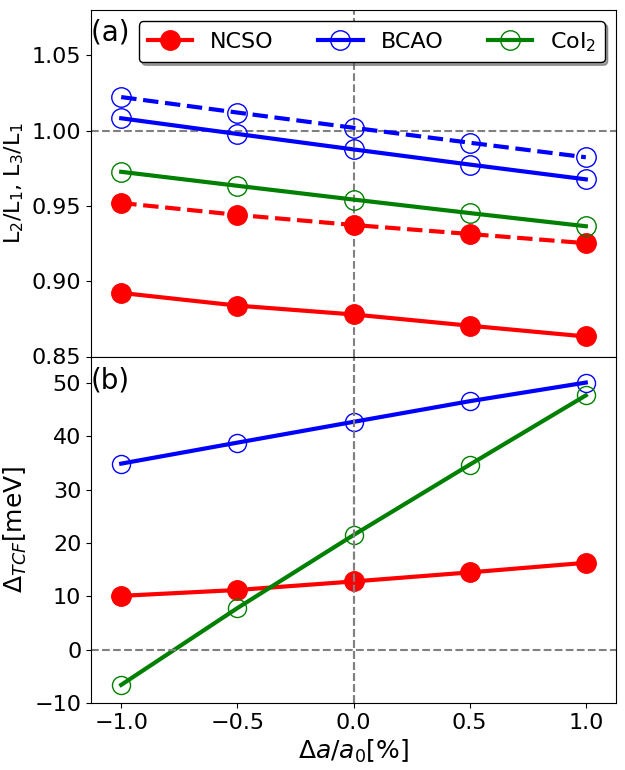}
\caption{(a) Trigonal distortions and (b) trigonal crystal field $\Delta_{TCF}$ versus strain of NCSO ($spd$-model), BCAO ($spd$-model), and CoI$_2$ ($pd$-model)}   
  \label{TCF of NCSO, BCAO, and CoI2}
\end{figure}

\section{DMFT Calculations for NCSO at $\Delta a/a_0=\pm1\%$}

\begin{figure}[htb!] 
   \includegraphics[width=0.44\textwidth]{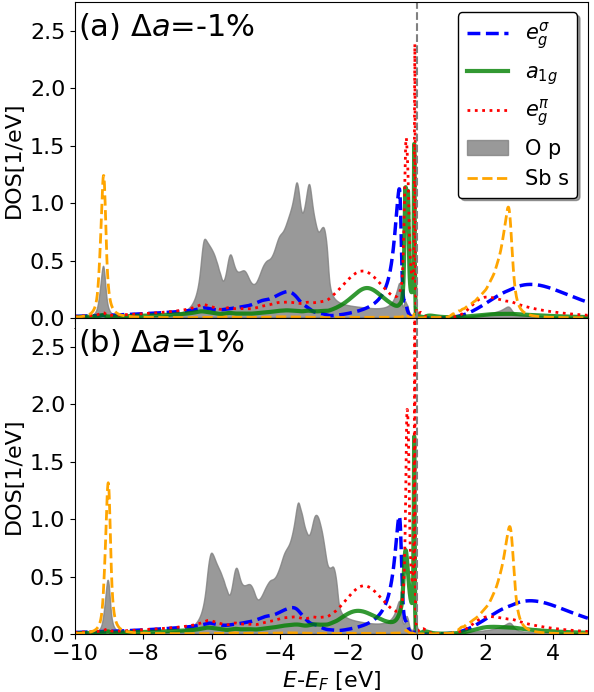}

   \hspace{0.1cm}
   
   \includegraphics[width=0.44\textwidth]{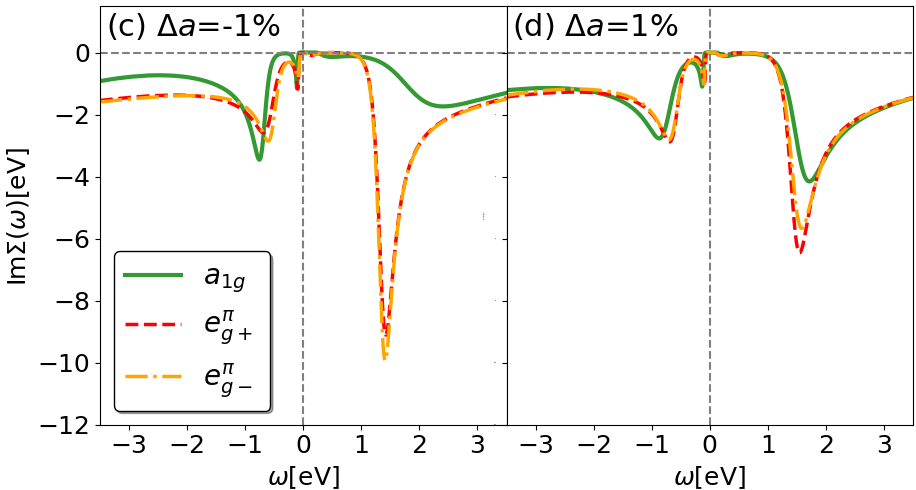}
\caption{DFT+DMFT DOS of NCSO with $U=5$ eV and $J=0.8$ eV for (a) compressive strain $\Delta a/a_0=-1\%$ and (b) tensile strain $\Delta a/a_0=1\%$. The corresponding DMFT self-energy $Im\Sigma(\omega)$ of $a_{1g}$, $e_{g+}^{\pi}$, and $e_{g-}^{\pi}$ orbitals for (c) compressive ($\Delta a/a_0=-1\%$) and (d) tensile ($\Delta a/a_0=1\%$) strains.} 
   \label{DMFT_Compress_Elongate}
\end{figure}

Here, we show DMFT calculations for NCSO at $\Delta a/a_0=\pm 1\%$. Under the compressive strain $\Delta a/a_0=-1\%$, the $a_{1g}$ orbital occupancy is almost fully filled ($N\sim 1.8$) consistently with the DMFT DOS (Fig. \ref{DMFT_Compress_Elongate}a). On the other hand, a noticeable unoccupied states of the $a_{1g}$ orbital is located at $E-E_F=2$ eV for the tensile strain $\Delta a/a_0=1\%$ (Fig. \ref{DMFT_Compress_Elongate}b), accompanied by a larger peaks of the imaginary self-energy of this orbital at $\omega\sim 2$ eV (Fig. \ref{DMFT_Compress_Elongate}d), indicating this orbital becomes more correlated in the tensile region. However, the electron correlation of the $a_{1g}$ orbital is still slightly smaller than the ones of the $e_{g}^{\pi}$ orbitals since the peak of the $a_{1g}$ imaginary self-energy is smaller than the ones of the $e_{g}^{\pi}$ imaginary self-energy at $\omega\sim1.5-2$ eV. Also, a slightly smaller (larger) peak of the imaginary self-energy of the $e_{g-}^{\pi}$ orbital than the one of the $e_{g+}^{\pi}$ orbital indicates a JT compression (elongation), consistent with the JT ratio (Co-O)$_{\hat{z}}$/(Co-O)$_{\hat{x}/\hat{y}}<1(>1)$ at $\Delta a/a_0=1\% (-1\%)$.

\section{DMFT Calculation for CoI$_2$}
\label{DMFT CoI2}

\begin{figure}[htb!]
\includegraphics[width=0.40\textwidth]{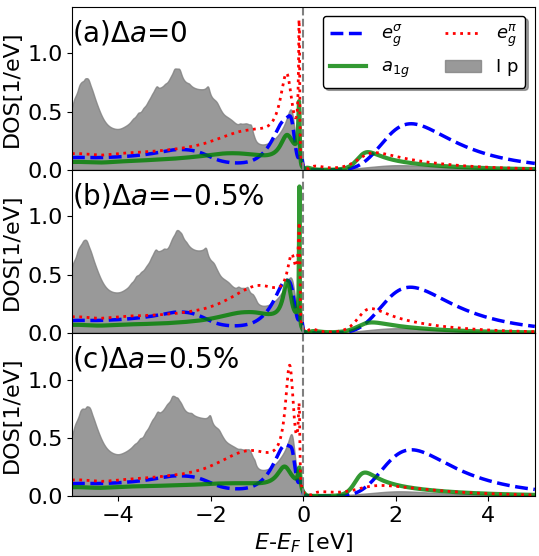}

\vspace{2mm}

\includegraphics[width=0.44\textwidth]{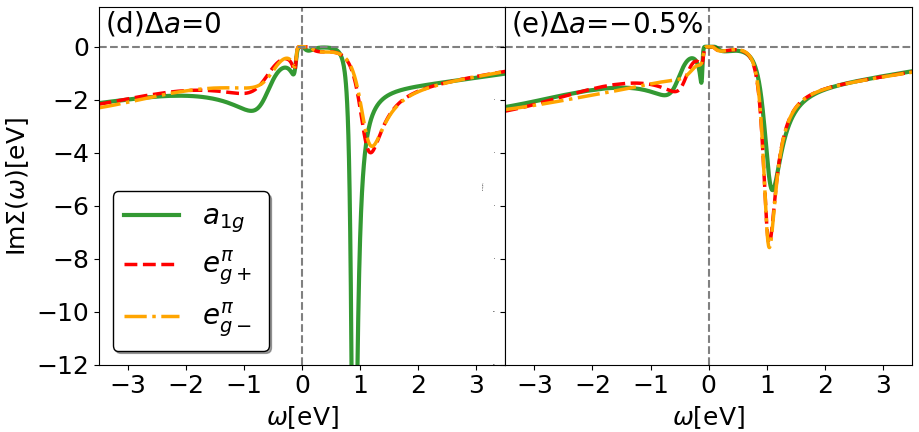}
\caption{(a,b,c) DFT+DMFT DOS of CoI$_2$ with $U=5$ eV and $J=0.8$ eV for $\Delta a/a_0=0,\pm 0.5\%$. (d,e) The corresponding self-energy $Im\Sigma(\omega)$ for $a_{1g}$, $e_{g+}^{\pi}$, and $e_{g-}^{\pi}$ orbitals of CoI$_2$ for $\Delta a/a_0=0$ and $-0.5\%$} 
   \label{CoI2 DMFT}
\end{figure}

Here, we show some DMFT results for CoI$_2$ at $\Delta a/a_0=0,\pm 0.5\%$. Fig. \ref{CoI2 DMFT}(a,b,c) show that the unoccupied states of $a_{1g}$ orbital increases as the strain effect becomes more tensile, indicating the $a_{1g}$ orbital becomes more correlated as the occupancy gets closer to the half filling. For the compressive strain $\Delta a/a_0=-0.5\%$, the imaginary self-energy of the $e_{g}^{\pi}$ orbitals has a stronger peak than the one of the $a_{1g}$ orbital at $\omega\sim 1$ eV, implying the $e_{g}^{\pi}$ orbitals are more correlated than the $a_{1g}$ one. 

\vspace{-8mm}

\section{Spin Exchange Interactions}
\label{Exchange Interacions}

Here, we show the spin exchange interactions obtained via the fourth-order strong coupling perturbation theory developed by Liu and Kee analysis \cite{PhysRevB.107.054420}. The pseudospin-1/2 interactions of the Kitaev Hamiltonian on $Z$-type nearest-neighbor (NN) bond have a general form~\cite{PhysRevLett.125.047201,Explanation}
\begin{align}
\mathcal{H}_{ij}^{(z)}=&J_K\Tilde{S}_i^z\Tilde{S}_j^z+J_H\Tilde{\bm{S}}_i\cdot\Tilde{\bm{S}}_j+\Gamma(\Tilde{S}_i^{x}\Tilde{S}_j^{y}+\Tilde{S}_i^{y}\Tilde{S}_j^{x})\notag
\\
&\Gamma'(\Tilde{S}_i^{x}\Tilde{S}_j^{z}+\Tilde{S}_i^{z}\Tilde{S}_j^{x}+\Tilde{S}_i^{y}\Tilde{S}_j^{z}+\Tilde{S}_i^{z}\Tilde{S}_j^{y})
\end{align}
where $J_K$ and $J_H$ are Kitaev and Heisenberg interactions respectively. The off-diagonal term $\Gamma$ arises from combined hoppings $t_1t_3$ and $t_2t_3$ where $t_1$, $t_2$, and $t_3$ are $d-d$ hoppings via the $t_{2g}$-$t_{2g}$ exchange channel~\cite{PhysRevB.107.054420,Winter_2022}. The $\Gamma'$ arises due to the trigonal crystal field that modifies the pseudospin $\Tilde{S}=1/2$ wave functions (Kramer doublet).

Although a finite $\Delta_{TCF}$ under the strain effect can affect $J_K$, $J_H$, $\Gamma$, and $\Gamma'$ in general~\cite{PhysRevLett.125.047201}, for simplicity we consider the ideal octahedron limit (no trigonal and JT distortions). 
The total Heisenberg $J_H$ and Kitatev $J_K$ interactions can be investigated via each anisotropic hopping channel:
\begin{align}
\label{Heiseneberg int.}
J_H&=J_{t_{2g}-t_{2g}}+J_{t_{2g}-e_{g}}+J_{e_{g}-e_{g}},
\\
\label{Kitaev int.}
J_K&=K_{t_{2g}-t_{2g}}+K_{t_{2g}-e_{g}}.
\end{align}
Within the intersite $U$, two-hole $2h$, and cyclic processes, each exchange interaction from the Eqs. (\ref{Heiseneberg int.}) and (\ref{Kitaev int.}) is given by

\begin{align}
J_{t_{2g}-t_{2g}}=&J_{t_{2g}-t_{2g}}^{U}
+J_{t_{2g}-t_{2g}}^{2h}+J_{t_{2g}-t_{2g}}^{cyclic}
\\
J_{t_{2g}-e_{g}}=&J_{t_{2g}-e_{g}}^{U}
+J_{t_{2g}-e_{g}}^{2h}+J_{t_{2g}-e_{g}}^{cyclic}
\\
J_{e_{g}-e_{g}}=&J_{e_{g}-e_{g}}^{U}
+J_{e_{g}-e_{g}}^{2h}
\\
K_{t_{2g}-t_{2g}}=&K_{t_{2g}-t_{2g}}^{U}
+K_{t_{2g}-t_{2g}}^{2h}+K_{t_{2g}-t_{2g}}^{cyclic}
\\
K_{t_{2g}-e_{g}}=&K_{t_{2g}-e_{g}}^{U}
+K_{t_{2g}-e_{g}}^{2h}+K_{t_{2g}-e_{g}}^{cyclic}
\end{align}

\begin{figure}[htb!]
  \centering
  \includegraphics[width=0.4\textwidth]{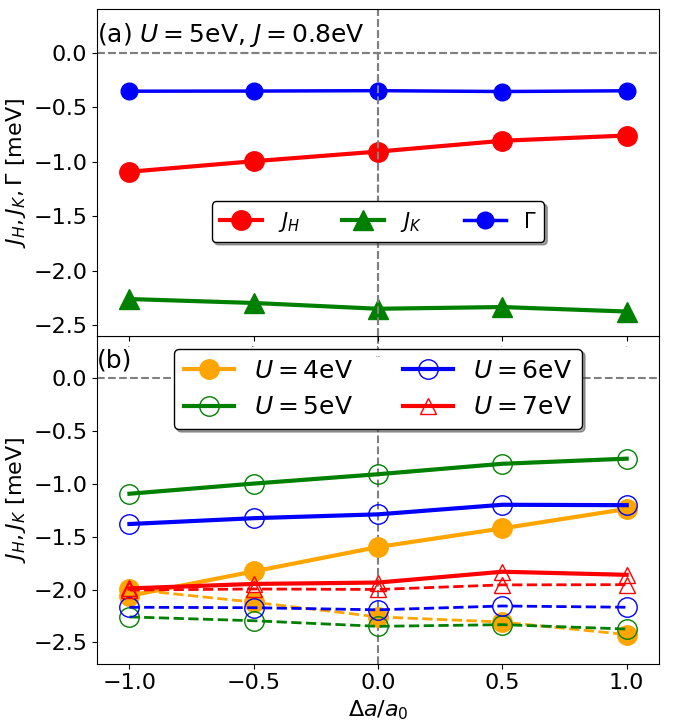}
\caption{(a) Heisenberg $J_H$,  Kitaev $J_K$, and off-diagonal $\Gamma$ interactions versus strain with $U=5$ eV and $J=0.8$ eV, and (b) Heisenberg (solid) and Kitaev (dashed) interactions versus strain with $J=$0.8 eV and various values of $U$. The exchange interactions are calculated using Liu and Kee analysis \cite{PhysRevB.107.054420} with $U_p$=0.7$U$ \cite{QSL_Condition_1} and $J_p$=0.3$U_p$ \cite{JpUp}. }   
  \label{J=0.8eV, vary U}
\end{figure}  

\begin{widetext}
Following the analysis of Liu and Kee~\cite{PhysRevB.107.054420}, the analytic expressions of these terms are given by
\begin{align}
\label{J_t2g_t2g_U}
J_{t_{2g}-t_{2g}}^{U}=&\frac{1}{486}\bigg(-\frac{171}{U-3J}+\frac{259}{U+J}+\frac{44}{U+4J}\bigg)t_1^2
+\frac{1}{54}\bigg(-\frac{21}{U-3J}+\frac{29}{U+J}+\frac{4}{U+4J}\bigg)t_2^2\notag
\\
+&\frac{2}{243}\bigg(-\frac{27}{U-3J}+\frac{43}{U+J}+\frac{8}{U+4J}\bigg)t_3^2
+\frac{4}{243}\bigg(\frac{18}{U-3J}-\frac{8}{U+J}+\frac{5}{U+4J}\bigg)t_1t_3
\\
J_{t_{2g}-t_{2g}}^{2h}=&\bigg[-\frac{80}{81}\frac{1}{2\Delta_{pd}+U_p-3J_p}+\frac{304}{243}\frac{1}{2\Delta_{pd}+U_p-J_p}+\frac{32}{243}\frac{1}{2\Delta_{pd}+U_p+2J_p}\bigg]\frac{(t_{pd}^{\pi})^4}{(\Delta_{pd})^2}
\\
J_{t_{2g}-t_{2g}}^{cyclic}=&\frac{2}{81\Delta_{pd}}\frac{(t_{pd}^{\pi})^4}{(\Delta_{pd})^2}
\end{align}

\begin{align}
J_{t_{2g}-e_{g}}^{U}=&\frac{5}{243}\bigg(-\frac{27}{U-3J+10Dq}+\frac{43}{U+J+10Dq}+\frac{8}{U+4J+10Dq}+\frac{24}{U+2J-10Dq}\bigg)t_6^2
\\
J_{t_{2g}-e_{g}}^{2h}=&\bigg[-\frac{10}{27}\frac{1}{2\Delta_{pd}+U_p-3J_p+10Dq}+\frac{250}{243}\frac{1}{2\Delta_{pd}+U_p-J_p+10Dq}+\frac{80}{243}\frac{1}{2\Delta_{pd}+U_p+2J_p+10Dq}\bigg]
\notag\\
&\times(t_{pd}^{\pi})^2(t_{pd}^{\sigma})^2\bigg(\frac{1}{\Delta_{pd}}+\frac{1}{\Delta_{pd}+10Dq}\bigg)^2
\\
J_{t_{2g}-e_{g}}^{cyclic}=&-\frac{40}{81}\frac{1}{2\Delta_{pd}+10Dq}(t_{pd}^{\pi})^2(t_{pd}^{\sigma})^2\bigg(\frac{1}{\Delta_{pd}}+\frac{1}{\Delta_{pd}+10Dq}\bigg)^2
\end{align}
\begin{align}
J_{e_{g}-e_{g}}^{U}=&\frac{100}{81}\frac{1}{U+2J}(t_4^2+t_5^2)
\\
J_{e_{g}-e_{g}}^{2h}=&\bigg[-\frac{200}{81}\frac{1}{2\Delta_{pd}+U_p-3J_p+2(10Dq)}+\frac{200}{81}\frac{1}{2\Delta_{pd}+U_p-J_p+2(10Dq)}\bigg]\frac{(t_{pd}^{\sigma})^4}{(\Delta_{pd}+10Dq)^2}
\end{align}
\begin{align}
\label{K_t2g_t2g_U}
K_{t_{2g}-t_{2g}}^{U}=&\frac{1}{243}\bigg(\frac{45}{U-3J}+\frac{11}{U+J}+\frac{28}{U+4J}\bigg)t_1^2
+\frac{1}{243}\bigg(-\frac{81}{U-3J}+\frac{73}{U+J}-\frac{4}{U+4J}\bigg)t_2^2\notag
\\
+&\frac{2}{81}\bigg(\frac{3}{U-3J}-\frac{7}{U+J}-\frac{2}{U+4J}\bigg)t_3^2
+\frac{1}{243}\bigg(-\frac{63}{U-3J}+\frac{31}{U+J}-\frac{16}{U+4J}\bigg)t_1t_3
\\
K_{t_{2g}-t_{2g}}^{2h}=&\bigg[\frac{40}{81}\frac{1}{2\Delta_{pd}+U_p-3J_p}-\frac{56}{243}\frac{1}{2\Delta_{pd}+U_p-J_p}+\frac{32}{243}\frac{1}{2\Delta_{pd}+U_p+2J_p}\bigg]\frac{(t_{pd}^{\pi})^4}{(\Delta_{pd})^2}
\\
K_{t_{2g}-t_{2g}}^{cyclic}=&-\frac{20}{81\Delta_{pd}}\frac{(t_{pd}^{\pi})^4}{(\Delta_{pd})^2}
\end{align}
\begin{align}
K_{t_{2g}-e_{g}}^{U}=&\frac{5}{243}\bigg(-\frac{9}{U-3J+10Dq}+\frac{1}{U+J+10Dq}-\frac{4}{U+4J+10Dq}-\frac{12}{U+2J-10Dq}\bigg)t_6^2
\\
K_{t_{2g}-e_{g}}^{2h}=&\bigg[-\frac{10}{81}\frac{1}{2\Delta_{pd}+U_p-3J_p+10Dq}-\frac{50}{243}\frac{1}{2\Delta_{pd}+U_p-J_p+10Dq}-\frac{40}{243}\frac{1}{2\Delta_{pd}+U_p+2J_p+10Dq}\bigg]
\notag\\
&\times(t_{pd}^{\pi})^2(t_{pd}^{\sigma})^2\bigg(\frac{1}{\Delta_{pd}}+\frac{1}{\Delta_{pd}+10Dq}\bigg)^2
\\
K_{t_{2g}-e_{g}}^{cyclic}=&\frac{20}{81}\frac{1}{2\Delta_{pd}+10Dq}(t_{pd}^{\pi})^2(t_{pd}^{\sigma})^2\bigg(\frac{1}{\Delta_{pd}}+\frac{1}{\Delta_{pd}+10Dq}\bigg)^2.
\end{align}
Also, the off-diagonal term $\Gamma$ is given by 
\begin{align}
\Gamma=\frac{4}{81}\bigg(\frac{3}{U-3J}-\frac{7}{U+J}-\frac{2}{U+4J}\bigg)t_1t_2+\frac{1}{243}\bigg(-\frac{63}{U-3J}+\frac{31}{U+J}-\frac{16}{U+4J}\bigg)t_2t_3.
\end{align}
\end{widetext}


\begin{figure*}[htb!]
  \centering
  \includegraphics[width=0.9\textwidth]{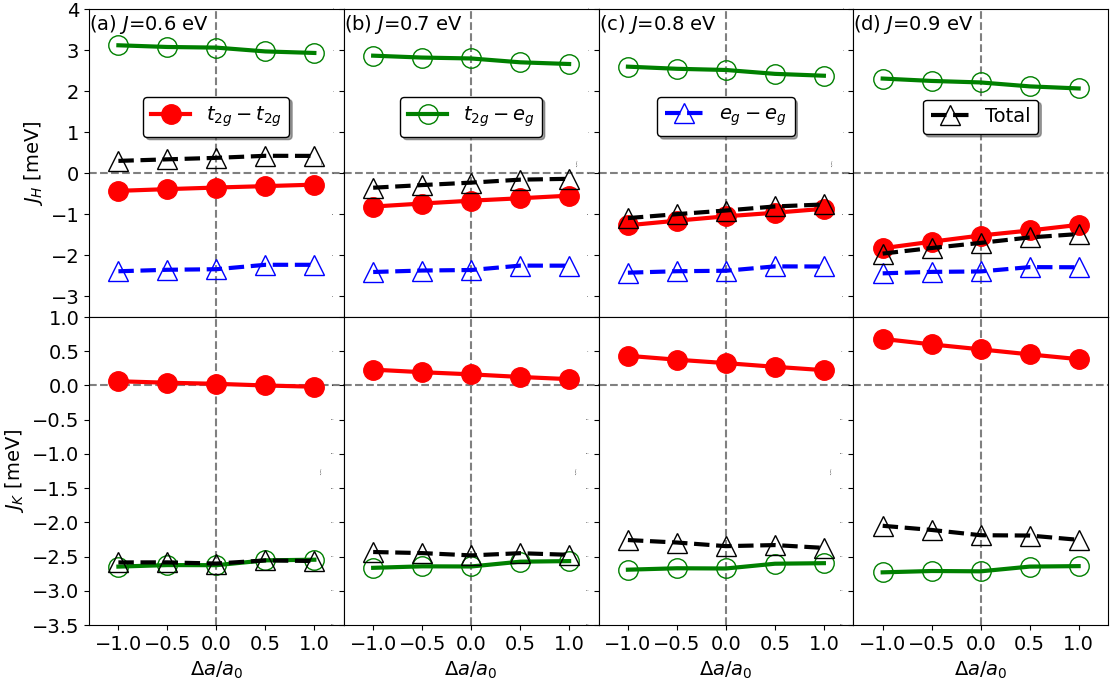}
\caption{Heisenberg and Kitaev interactions from each exchange channel and their total versus strain calculated using Liu and Kee analysis \cite{PhysRevB.107.054420} with $U=5$ eV, $U_p$=0.7$U$ \cite{QSL_Condition_1}, and $J_p$=0.3$U_p$ \cite{JpUp} while varying the Hund's coupling $J$. }   
  \label{exchange channels}
\end{figure*}

Using the above equations, the magnetic exchange interactions $J_H$, $J_K$, and $\Gamma$ under strain effect can be obtained with our material parameters obtained from DFT. For $U=5$ eV and $J=0.8$ eV, we found that the Kitaev $|J_K|$ term is the largest while the Heisenberg $|J_H|$ and the off-diagonal $|\Gamma|$ terms are still finite at zero strain. Within the limit of perfect octahedron, $|J_K|$ and $|\Gamma|$ are not sensitive to strain, but $|J_H|$ is reduced under the tensile strain (Fig. \ref{J=0.8eV, vary U}a) due to the reducing of $|t_3|$ hopping (Fig. \ref{hopping_charge_transfer}a). Our $J_K$ and $J_H$ results at zero strain are consistent with those obtained from the DFT spin-exchange calculation~\cite{INS_TCF} and the trends of these values under strains are also
similar to those in Cu$_{3}$Co$_{2}$SbO$_6$~\cite{XAS_TCF}.

We also found that the Hubbard $U$ and Hund's coupling $J$ values can also affect the exchange interactions. While fixing $J$=0.8 eV and varying $U$, we found that the system is pushed closest to the QSL ($J_H\sim 0$) at $U=5$ eV, but the $J_H$ dependence is not much sensitive to the strain, compared to the one at $U=4$ eV (Fig. \ref{J=0.8eV, vary U}b). On the other hand, while fixing $U=5$ eV and varying $J$, we found that the overall Heisenber interaction $J_H$ is very sensitive to $J$ (Fig. \ref{exchange channels}). For $J=0.6-0.7$ eV ($U=5$ eV) the overall Heisenberg interaction $J_H$ is small, while the FM Kitaev one is large due to the dominant contribution from the $t_{2g}$-$e_{g}$ channel, consistent with Liu $et$ $al$ analysis for $d^7$ Co compounds \cite{PhysRevLett.125.047201}. 
We found that the $J_H$ generated from the $e_g$-$e_g$ channel is not very sensitive to the strain and also remains unchanged as $J$ varies, consistent with Liu and Khaliullin analysis \cite{QSL_Condition_1}. However, the AFM Heisenberg interaction from the $t_{2g}$-$e_{g}$ channel weakens the combined $t_{2g}$-$t_{2g}$ and $e_g$-$e_g$ FM Heisenberg interactions, thus reducing the overall Heisenberg interaction $J_H$.
As the Hund's coupling $J$ gets larger ($J\geq 0.8$ eV), the contribution from the $t_{2g}$-$t_{2g}$ channel gets larger, therefore, the total $J_H$ is enhanced while the total $J_K$ is weakened.

\end{document}